\documentclass[a4paper,fleqn,usenatbib]{mnras}

\usepackage{newtxtext,newtxmath}

\usepackage[T1]{fontenc}
\usepackage{ae,aecompl}

\usepackage{gensymb} 

\usepackage{graphicx}	
\usepackage{amsmath}	
\usepackage{amssymb}	

\usepackage[]{units} 

\usepackage{mathrsfs} 
\usepackage{amsbsy} 
\usepackage{color}
\usepackage{soul} 

\usepackage{tablefootnote}
\usepackage{epstopdf}

\title[Black hole reverberation]{Multi-timescale X-ray reverberation mapping of \\
	accreting black holes}

\author[Mastroserio et al.
]{
Guglielmo Mastroserio,$^{1}$\thanks{E-mail: g.mastroserio@uva.nl}
Adam Ingram$^{1,2} $ and
Michiel van der Klis$ ^{1} $
\\
$ ^{1} $Astronomical Institute, Anton Pannekoek, Univerity of Amsterdam, Science Park 904, NL-1098 XH Amsterdam, Netherlands\\
$ ^{2} $Department of Physics, Astrophysics University of Oxford, Denys Wilkinson Building, Keble Road, Oxford OX1 3RH, UK\\
}

\date{Accepted XXX. Received YYY; in original form ZZZ}

\pubyear{2017}

\begin{document}
\label{firstpage}
\pagerange{\pageref{firstpage}--\pageref{lastpage}}
\maketitle

\setstcolor{red}

\begin{abstract}
Accreting black holes show characteristic reflection features in their 
X-ray spectrum, including an iron K$\alpha$ line, resulting from 
hard X-ray continuum photons illuminating the accretion disk. 
The reverberation lag resulting from the path 
length difference between direct and reflected emission 
provides a powerful tool to probe the innermost regions around both stellar-mass and 
supermassive black holes.
Here, we present for the first time a reverberation mapping formalism that enables 
modeling of energy dependent time lags and variability amplitude for a wide 
range of variability timescales, taking the complete information of the 
cross-spectrum into account. 
We use a pivoting power-law model to account for the spectral variability of the 
continuum that dominates over the reverberation lags for longer time 
scale variability. 
We use an analytic approximation to self-consistently account for the non-linear 
effects caused by this continuum spectral variability, which have been ignored 
by all previous reverberation studies. 
We find that ignoring these non-linear effects can 
bias measurements of the reverberation lags, particularly at low frequencies. 
Since our model is analytic, we are able to fit simultaneously for a 
wide range of Fourier frequencies without prohibitive computational expense.
We also introduce a formalism of fitting to real and imaginary parts 
of our cross-spectrum statistic, which naturally avoids some 
mistakes/inaccuracies previously common in the literature.
We perform proof-of-principle fits to Rossi X-ray Timing Explorer data of Cygnus X-1.
 		
\end{abstract}




\section{Introduction}
\label{sec:intro}

In black hole X-ray binaries (BHBs) and active galactic nuclei (AGN), 
the central black hole is thought to be fed, at least in part, 
by an optically thick accretion disc that radiates a multi-temperature blackbody
spectrum \citep{Shakura1973}.
This disc emission peaks in soft X-rays for BHBs and optical soft X-rays for AGN. 
In both cases, a different component dominates the hard X-ray radiation, 
which is often described by a cut-off power-law. 
This \textit{continuum} emission is thought to 
originate from the Compton up-scattering of comparatively cool photons 
by hot electrons located in an optically thin ($\tau \sim 1$) region 
close to the black hole, often referred to as the corona \citep{Eardley1975,Thorne1975}.
Some fraction of these continuum photons illuminate the disc to be 
scattered into our line-of-sight, giving rise to a characteristic \textit{reflection} spectrum 
that imprints features onto the observed spectrum including a prominent iron K$\alpha$
fluorescence line at $\sim 6.4$ keV and a reflection hump peaking at $\sim 30 $ keV.
This reflection spectrum provides a powerful diagnostic for the dynamics of the 
accretion disc, since it is distorted by the gravitational pull of the black hole and 
Doppler shifts from rapid orbital motion \citep{Fabian1989}.
Rapid variability of the system provides another powerful diagnostic, particularly 
because any fluctuations in the continuum should be followed, after 
a light-crossing delay, by similar fluctuations in the reflection spectrum. 
Characterization of these reverberation lags provides another tool to map the 
accretion disc.

Reverberation lags can be probed by studying the Fourier frequency dependent time 
lags between different energy bands, since bands with a greater contribution from reflection
should slightly lag those dominated by the continuum.
The time lags can be calculated from the argument of the cross-spectrum between 
each energy channel and a common reference band \citep{Uttley2014}. 
It has long been known that hard photons lag soft photons for comparatively 
low Fourier frequencies (below $\sim300 [M_\odot/M]\, {\rm Hz}$), both in BHBs 
\citep[e.g.][]{Miyamoto1989,Nowak1999} and AGN \citep[e.g.][]{Papadakis2001,McHardy2004}.
However, these lags do not show any reflection features in the lag-energy spectrum 
and for this reason they are thought to be associated with intrinsic variation of the continuum spectral shape. 
This is commonly interpreted as propagation of mass accretion rate fluctuations
towards the black hole on a viscous timescale \citep{Lyubarskii1997,Kotov2001,Arevalo2006,Ingram2013,Rapisarda2016}. 
This intrinsic continuum lag reduces with increasing Fourier frequency, leaving the opportunity 
to detect a reverberation signature at high frequencies. 
Such lags have been detected for AGN, first in the form of soft-excess emission 
($\sim 0.2-0.9$ keV) lagging the continuum dominated band ($\sim 1-4$ keV)
\citep[][]{Fabian2009}, and later in the form of an iron K feature in the lag-energy spectrum 
at $\sim 6.4$ keV \citep[][]{Zoghbi2012,Kara2016}. 
The latter is the cleanest measurement, since the proposed reflection origin of the soft X-ray 
excess in AGN is not universally accepted, with alternative models invoking an extra
Compton up-scattering component \citep[][]{Page2004,Done2012}, while 
it is very difficult to reproduce the iron K feature in the lag-energy spectrum without 
using the reflection mechanism.
Reverberation lags have not yet been clearly detected for BHBs 
\citep[even though][ recently found hints of FeK reverberation]{DeMarco2017}, 
since the smaller size of these systems leads to the lags being shorter ($\sim$ millisecond)  
and only dominant over the continuum lags for higher Fourier frequencies.
However, \citet{Uttley2011} and \citet{DeMarco2015,DeMarco2016} found that the disc 
blackbody emission lags the continuum emission in GX 339-4 and H1743-322 
by a few milliseconds,
which they attribute to reprocessed photons being re-emitted as thermalised radiation.

The reverberation signature also depends on Fourier frequency, since fast variations 
in the driving continuum are washed out in the reflected emission by path length 
differences between photons reflecting from different parts of the disc. 
This means that the iron K feature in the lag-energy spectrum should be 
broader at higher Fourier frequencies, since rapid variability is washed out for 
reflection from all but the smallest, most rapidly rotating and gravitationally redshifted disc radii.
Indeed, \citet{Zoghbi2012} found just this for the iron K lags in NGC 4151.
Further information is contained in the variability amplitude of the reflected 
emission relative to the continuum. 
\citet{Revnivtsev1999} and \citet{Gilfanov2000} found that the relative variability amplitude 
of the reflected emission in Cygnus X-1 decreases at higher Fourier frequencies, as expected
(see Section~\ref{sec:simulation}).

An elegant way to model reverberation is to calculate a \textit{response function}, 
defined as the energy and time dependent reflected emission resulting 
from a $\delta-$function flash in the driving continuum 
\citep[e.g.][]{Campana1995,Reynolds1999,Kotov2001}.
If the disc properties are approximately independent of the irradiating flux, 
the reflected flux responding to an arbitrary driving continuum signal is given by 
a convolution between the driving signal and the response function. 
The convolution theorem therefore means it is most convenient to consider the Fourier 
transform of the response function, referred to here as the 
\textit{transfer function} \citep[e.g.][]{Oppenheim1975}.
This function, which can in principle be constrained from 
energy and Fourier frequency dependent amplitude 
and phase of the observed cross-spectrum,
contains information about the accretion geometry.
However, the dominance of continuum lags at low frequencies makes this challenging.
Authors have therefore previously modelled the lags only at high Fourier frequencies 
\citep[e.g.][]{Cackett2014}, used an \textit{ad hoc} prescription to account for the 
continuum lags \citep[][]{Emmanoulopoulos2014}, or only considered amplitude 
and not phase \citep[][]{Gilfanov2000}.  
Progress has been made in modelling the intrinsic continuum lag with propagating 
fluctuations and taking account of reverberation lags in specific geometries 
\citep[][]{Wilkins2013,Wilkins2016,Chainakun2017}. However this is computationally 
expensive, particularly for the purpose of fitting lag-energy spectra for a large 
range of Fourier frequencies. 
Here, we present a simple analytic way  to model the continuum lags, 
and self-consistently take into account the impact of those lags on the 
reverberation signal.
We model the continuum lags as perturbations in the continuum power-law 
index and account for the changes in the reflection spectrum caused by these perturbations
with a 1st order Taylor expansion.
A similar pivoting power-law model was considered by \citet[][]{Poutanen2002}, 
but he assumed the energy dependent reflection to continuum ratio to 
be \textit{independent} of the illuminated continuum. 
Our formalism allows us to fit to data, considering lags and amplitude for a 
large range of Fourier frequencies without prohibitive computational cost. 

In Section~\ref{sec:cross} we introduce the cross-spectral method that we 
use to compute the model and analyse the data. 
We highlight some common inaccuracies that can occur with similar techniques,
and show that such inaccuracies can be easily avoided by modelling
real and imaginary parts of the cross-spectrum rather than the amplitude
and phase. 
In Section~\ref{sec:model}, we present our model formalism. 
In Section~\ref{sec:simulation}, we explore our model parameters, focusing 
on the importance of the non-linear effects resulting from variations 
in the energy dependence of the reflection disc caused by variations 
in the hardness of the driving continuum.
In Section~\ref{sec:fit_data}, we perform proof of principle fits 
to Cygnus X-1 data.

\section{Cross-Spectrum Method}
\label{sec:cross} 
This Section briefly reviews the spectral timing 
techniques previously used, before describing our technique 
that allows amplitude and phase as a function of
energy and frequency to be modelled. 
We also describe how our method naturally corrects some 
mathematical inaccuracies often encountered in the literature.

First we define a set of complex cross-spectra $\langle
C\left(E,\nu\right) \rangle$
as a function of energy and frequency
\begin{equation}
\langle C\left(E,\nu\right)\rangle = \langle S\left(E,\nu\right)\,F^*\left(\nu\right)\rangle,
\label{eq:cross-spectrum}
\end{equation} 
where $S\left(E,\nu\right)$ is a set of Fourier transforms of the signal light curve in different 
energies $E$, and $F\left(\nu\right)$ is the Fourier transform 
of the signal in an arbitrary reference band.  
Starred quantities and angle brackets denote the complex conjugate and 
ensemble averaging respectively. The phase lag for each energy band
relative to the reference band is
\begin{equation}
\phi(E,\nu) = \arg\left[\langle C\left(E,\nu\right)\rangle\right].
\label{eq:lag}
\end{equation}
Many works have focused on analysing the phase-lag with
reverberation models 
\citep[][]{Kotov2001,Poutanen2002,Zoghbi2011,Emmanoulopoulos2014,Wilkins2016,Chainakun2017}. 
Although these studies constrain model parameters by fitting either
lag-frequency or lag-energy spectra (sometimes both), they all neglect the
information included in the cross-amplitude.

With a suitable choice of normalization, the variability amplitude 
(in units of absolute rms) as a function of energy and frequency,
$\sqrt{\langle|S\left(E,\nu\right)|^2\rangle}$, can be calculated directly
\citep{Revnivtsev1999} by measuring the power spectrum 
averaged over the frequency range $\Delta$ for each energy channel
\begin{equation}
\sqrt{\langle|S\left(E,\nu\right)|^2\rangle} = \sqrt{\left[\langle P\left(E,\nu\right)\rangle -P_{\rm noise}\left(E\right)\right]\,\Delta}
\label{eq:ampl}
\end{equation} 
where the $P\left(E,\nu\right)$ and $P_{\rm noise}\left(E\right)$ are
respectively the power spectra and Poisson noise measured for each
energy channel \citep[see e.g.][]{Klis1989,Uttley2014}. 
The correlated variability amplitude, a related quantity, can be calculated with higher 
signal-to-noise using the covariance spectrum 
\citep{Wilkinson2009,Uttley2014}
\begin{equation}
| \langle G(E,\nu) \rangle | = \frac{  \sqrt{\Delta}\, | \langle
	C\left(E,\nu\right) \rangle | } { \sqrt{\langle P(\nu)\rangle -P_{\rm noise}} } = \gamma_{\rm c}(E,\nu) \sqrt{\langle|S\left(E,\nu\right)|^2\rangle},
\label{eq:covariance}
\end{equation}
where $\gamma_{\rm c}(E,\nu)$ is the coherence between each energy channel and
the reference band, and $\langle P(\nu) \rangle $ and $P_{\rm noise}$ are respectively
the power spectrum and Poisson noise contribution for the reference band 
(in units of absolute rms squared per Hz)
\footnote{$P_{\rm noise}$ does not have the angle brackets 
		because it is estimated theoretically from the assumption of 
		pure Poissonian counting noise.}.
So, the covariance is the correlated variability
amplitude, which is related to the variability amplitude through the
coherence function \citep{Vaughan1997}. 
Since the coherence function is often close to unity for accreting compact objects
\citep[e.g.][]{Nowak1999}, the covariance gives a good estimate of
the variability amplitude, and its error bars are smaller if a high
count rate reference band is chosen \citep{Wilkinson2009}.

Many authors have modelled the variability amplitude as a function of
energy and frequency \citep[\textit{frequency resolved spectroscopy}:
e.g.][]{Gilfanov2000,Axelsson2013},
either using the power spectrum or the covariance. 
This approach, as with the lag modelling, 
provides strong constraints, but now neglects the information
contained in the phase lags. 
Although past works have discussed both
amplitude and phase together in the context of reverberation
\citep[e.g. ][]{Uttley2011,Kara2013a},
none so far have used quantitative fitting of models for 
the energy and frequency dependent amplitude and phase to data.

We consider both the amplitude and phase jointly 
by considering the \textit{complex covariance}, defined as
\begin{equation}
\langle G\left(E,\nu\right) \rangle= \frac{\sqrt{\Delta}\, \langle C(E,\nu) \rangle}{\sqrt{\langle P\left(\nu\right)\rangle-P_{\rm noise}}}.
\label{eq:complex_covariance}
\end{equation} 
We fit models to data for the real and imaginary parts of $\langle
G\left(E,\nu\right) \rangle$ as a function of energy, for a
number of discrete frequency ranges \citep[following][]{Rapisarda2016,Ingram2016}. 
Fitting for real and imaginary parts rather than amplitude and phase naturally avoids some
mistakes and inaccuracies commonly found in the literature. 
This is because the linearity inherent in the Fourier transform operation is preserved. 
For instance, in order to fit to data, the model must be adjusted for the instrument response. 
This is a trivial operation in our case, involving simply convolving 
real and imaginary parts of the model complex covariance 
with the instrument response, meaning that the model can simply 
be loaded into e.g. \textsc{xspec} as if it were a spectral model
(in order to do this correctly it is important to choose a normalization such that the 
modulus of the Fourier coefficients is in units of absolute rms). 
For the amplitude and phase it is not possible to apply the same procedure. 
In particular with the amplitude, it has become commonplace 
in the literature to account for the instrument response
by convolving the model for $\sqrt{\langle|S\left(E,\nu\right)|^2\rangle}$
with the instrument response.
We show in Appendix~\ref{app:cross-spectrum_analysis} 
that this is mathematically incorrect, 
unless $\phi(E,\nu)=0$ \citep[which is often approximately true, 
but non-zero phase lags are of physical interest and can in practice 
be as large as $180^{\circ}$ e.g. Cygnus X-2:][]{Mitsuda1989}.
If there would be a need to fit directly for phase and/or amplitude, the 
correct procedure would be to convolve real and imaginary parts of the 
complex covariance (or cross-spectrum) with the instrument response and \textit{then} compare
the modulus and the argument of this `folded' complex covariance 
(or cross-spectrum) to amplitude and phase measured from the observed data.

Another advantage of our method is that $\langle G\left(E,\nu\right)\rangle$ 
can be easily modelled as the sum of multiple spectral components. 
Although it has become commonplace in the
literature to model $\langle |S\left(E,\nu\right)| \rangle$ as a sum
of components, this is not mathematically correct in general, since
the components should really be complex quantities summed as vectors
on the complex plane (see Appendix~\ref{app:cross-spectrum_analysis}). 
Therefore summing spectral components of the amplitude is
only appropriate if the phase difference between all complex
components is zero. 
\citet{Kotov2001} point out that the error is small if one component 
is small compared with the other, but this is often not the case. 
Similarly, multiple components are often required to model the observed lags. 
For example, one component may contribute the continuum
lags and the other the reverberation lags. 
We note that it is mathematically incorrect to simply add the model continuum lag to the
model reverberation lag, even in a small angle approximation. 
There are many instances in the literature where it is not clear whether or
not this mistake has been made \citep[e.g.][]{Poutanen2002,Zoghbi2011,Emmanoulopoulos2014}. 
These difficulties are naturally avoided for our method \citep[see also][]{Rapisarda2016,Ingram2016}. 


\section{Model Formalism}
\label{sec:model} 

We consider emission from two main components: a continuum cut-off 
power-law spectrum emitted by a point-like source 
(this could approximately represent a spectrum due to inverse-Compton scattering), 
and the same radiation reflected from the disc
(or rather scattered in the disc atmosphere).
We do not consider intrinsic thermal disc emission, under the assumption that this peaks 
outside of our considered energy range (> 3 keV). 
We first describe the two components separately and then we consider them 
together to probe the non-linear effects resulting from variability of the
continuum shape.  
In this section we define the theoretical model we use to describe the data. 
	Therefore we drop the angle brackets and we can write 
	$ |S\left(E,\nu\right)| $ as a prediction of $ \sqrt{\langle|S\left(E,\nu\right)|^2\rangle} $.
In the whole paper $E$ refers to the photon energy seen by the observer and 
$E_{\rm em}$ to the energy emitted in the local frame co-moving with the disc plasma.

\subsection{Continuum}
\label{sub_sec:continuum} 
The continuum emission is due to photons that are Compton up-scattered by hot electrons.
In the case of the continuum emission we consider $E = E_{\rm em}$. 
This is a relatively crude approximation that is routinely made
in the literature, since the continuum is almost featureless (apart from the high energy cut-off) 
and therefore the energy shifts are less important than the reflection spectrum
\citep[although see][for a discussion on this point]{Niedzwiecki2016}.
The spectrum of this emission can be described as a power-law, 
cut off at energy $E_{\textmd{cut}}$, that varies in time as
\begin{equation}
	D\left(E,t\right) = A\left(t\right)E^{-\Gamma+\beta\left(t\right)}{\rm e}^{-\nicefrac{E}{E_{\textmd{cut}}}}.
	\label{eq:continuum_non-lin}
\end{equation}
$D$  expresses the specific photon flux of the continuum at the observer,
and we consider the normalization $A\left(t\right)$  and the power-law index 
$-\Gamma+ \beta\left(t\right) $ to both be time dependent.
$E_{\rm cut}$ could also be variable in general, but for 
simplicity we assume this to be constant. 
We also note that Eq.~\ref{eq:continuum_non-lin} differs somewhat at 
high energies from the true shape of a spectrum generated 
through Compton up-scattering of photons by a thermal 
population of electrons \citep[e.g.][]{Zdziarski2003}, 
but will suffice for our purposes.

Fluctuations in the normalization and the power-law index could be due to, e.g. fluctuations
of the mass accretion rate in the disc, or variation in the temperature of the 
Comptonising region.
Another way to produce these fluctuations is through 
the rising of compact magnetic flares from the accretion disc
\citep{Poutanen1999}, although the observed linear rms-flux 
relation rules out a simple model in which these flares are 
statistically independent \citep{Uttley2005}. 

Here, we simply use Eq.~\ref{eq:continuum_non-lin}
as a mathematical model to describe 
what varies with time in the continuum emission. 
This model was briefly considered 
by \citet{Kotov2001} who noted the observed phase lags 
(in Cygnus X-1) are not consistent with a simple scenario in which
$\beta(t) \propto A(t)$; instead there must be a delay 
between $\beta(t)$ and $A(t)$ oscillations and this was 
explored in more detail by \citet{Kording2004}
\citep[also see][for similar models applied with different purpose]{Shaposhnikov2012,Misra2013}.	

We use the lamppost geometry \citep{Matt1992} in which the continuum is emitted
isotropically by a point source situated 
on the black hole spin axis and with a 
stationary, cylindrically symmetric thin 
prograde disc in the equatorial plane.
This drastically simplifies our calculations.

		\subsection{Reverberation}
		 \label{sub_sec:reflection} 
		Some fraction of the continuum photons are reflected from the disc into our line
		of sight. Suppose the ionization structure of
		the disc does not change much on short timescales. Then the reflection
		spectrum observed from a patch of the disc of area subtending a solid 
		angle according to the observer of ${\rm d}\Omega \left(r,\phi\right)$, 
		where coordinates~$(r,\phi)$ are  
		respectively disc radius and azimuth, is
		\begin{align}
		dR(E,t|r,\phi) = &\varepsilon\left(r\right) g^3\left(r,\phi\right) A{\big (} t -\tau(r,\phi){\big )}\nonumber\\
		&\mathscr{R}{\big (} E/g(r,\phi)|\Gamma-\beta\left(t-\tau(r,\phi)\right){\big )} {\rm d} \Omega .
		\label{eq:refl_patch}
		\end{align}
		The reflection energy spectrum varies with time because the incident continuum
		radiation does. 
		In Eq.~\ref{eq:refl_patch}, $\mathscr{R}\left(E/g|\Gamma-\beta\right)$ represents the plasma
		restframe reflection energy spectrum (in units of specific flux) emerging from the X-ray illuminated disc
		at coordinate $(r,\phi)$ and $g\left(r,\phi\right) \equiv E/E_{\rm em}$
		is the blue shift resulting from Doppler \& relativistic effects.
		The factor $g^3\left(r,\phi\right)$ accounts for Doppler boosting,  
		 and gravitational redshift of photons travelling from disc to observer,
		while $\varepsilon\left(r\right)$ is the geometrical 
		correction for the flux of photons travelling from source to disc 
		(both $g$ and $\varepsilon$ are defined in the Appendix~\ref{app:transfer_1}).
		In Eq.~\ref{eq:refl_patch}, and throughout this paper, we express 
		distances in units of $R_{\rm g}=GM/c^2$.
		The observed variations of the normalization and the power-law index 
		of the continuum radiation are delayed by an interval $\tau\left(r,\phi\right)$. 
		This is the time difference between the reflected and the direct 
		signals reaching the observer (see Appendix~\ref{app:transfer_1}).
		A given value of $\tau$ defines iso-delay curves on the disc that 
		the observer sees to be simultaneously illuminated 
		with the same continuum normalization $A$ and power-law index $-\Gamma + \beta$.
		This is the only correction we apply to the incident emission, 
		so we ignore that every radius of the disc sees 
		a different energy shift in the continuum energy spectrum due to
		gravitational redshift or blueshift, causing a different
		incident flux and power-law cut-off for every radius of the disc. 
		We use the model \textsc{xillver} \citep{Garcia2010,Garcia2013} to 
		calculate $\mathscr{R}$. 
		This model calculates the reflection spectrum by solving the 
		equations of radiative transfer, energy balance, 
		and ionization equilibrium in a Compton-thick, plane-parallel medium,
		being irradiated by a cut-off power-law spectrum.
		
\begin{figure}
	\includegraphics[width=\columnwidth]{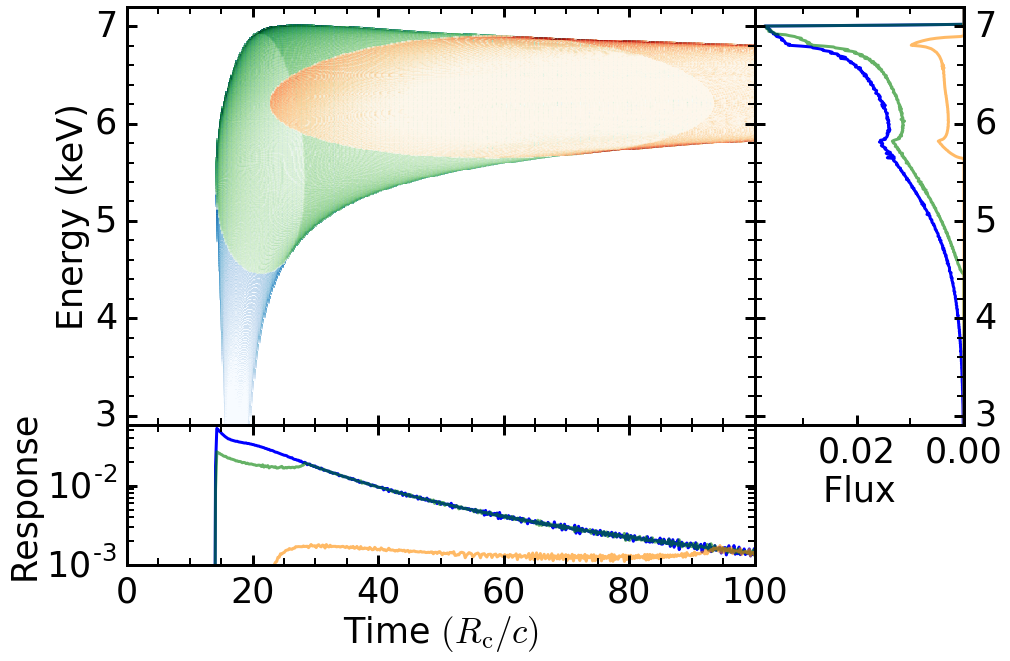}
	\caption{Central panel: Response function calculated for 
		three different values of $r_{\rm in}$ of the disc:
		$1.2\,R_{\rm g}$ (blue), $10\,R_{\rm g}$ (green), and $50\,R_{\rm g}$ (red). 
		For all three functions we consider a disc with $i=45\degree$ and 
		$r_{\textmd{out}}=10^6\,R_{\rm g}$ illuminated by a flash of emission 
		from a point like source at height $h=10\,r_{\rm g}$ above a black hole with spin $a=0.998$. 
		Here $\mathscr{R} = \delta$ ($E-6.4$ keV).  
		Right panel: Time integrated spectrum (i.e line profile). 
		Bottom panel: Energy integrated flux (i.e. impulse response function).}
	\label{fig:impulse_response}
\end{figure}

\begin{figure}
	\includegraphics[width=\columnwidth]{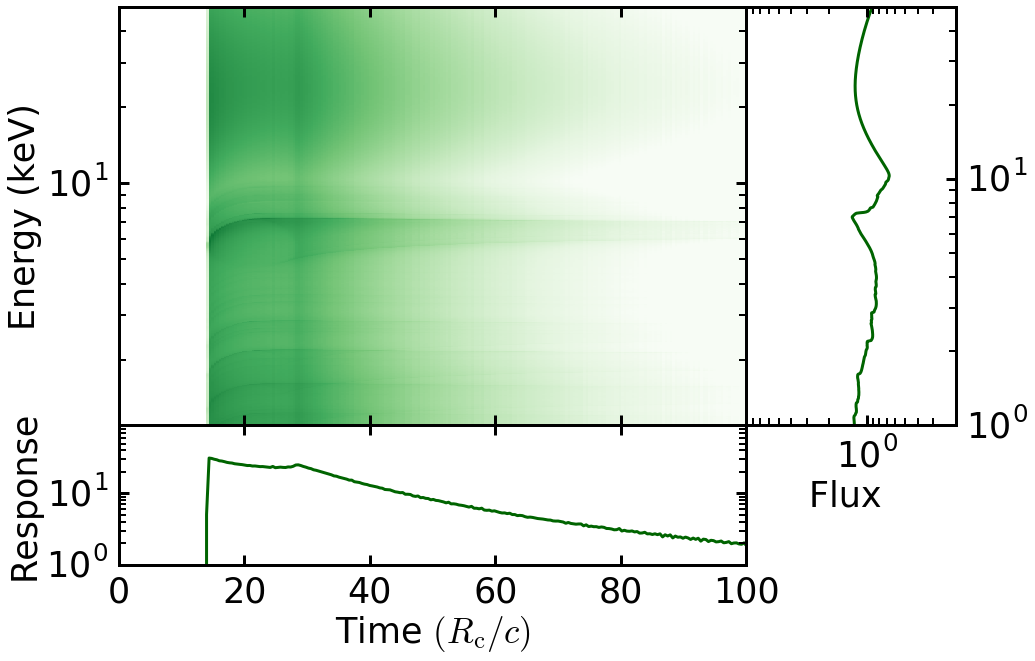}
	\caption{Central panel: Response function for a disc with $i=45\degree$,  
		$r_{\textmd{in}}=10\,r_{\rm g}$, and $r_{\textmd{out}}=10^6\,R_{\rm g}$ illuminated by a flash
		from a point like source at height $h=10\,R_{\rm g}$ above a black hole with spin $a=0.998$. 
		Here $\mathscr{R}$ is calculated using \textsc{xillver} with the parameters specified in the text.  
		Right panel: Time integrated spectrum (i.e. energy spectrum). 
		Bottom panel: energy integrated flux (i.e. impulse-response function).}
	\label{fig:impulse_response_xillver}
\end{figure}

		The observed reflection spectrum can be calculated by integrating 
		Eq.~\ref{eq:refl_patch} over the entire disc surface. 
		This can be simplified greatly by ignoring variations in the 
		power-law index, i.e. setting $\beta=0$. 
		In this case, the variations in both the continuum and
		reflected radiation are \textit{linear} and we can therefore calculate
		the observed reflection spectrum time series by convolving the 
		restframe spectrum time series with the response function. 
		In this case, we can write
		\begin{equation}
		R\left(E,t\right) = 
		A\left(t\right)\otimes w\left(E ,t \right),
		\label{eq:reflection_tot_conv}
		\end{equation}
		where the operation $\otimes$ denotes a convolution in the time domain
		(see Appendix~\ref{app:linearization} for the definition) and 
		 $w(E,t)$ is the response function. 
		 Here, we use a simplified calculation of the response function employing  
		 the Kerr metric to calculate the energy shift as a function of $r$ and
		 $\phi$, but using a flat spacetime for calculating the light travel time 
		 of both continuum and reflection photons. 
		 In this case, ${\rm d}\Omega = r\, {\rm d}r\,{\rm d}\phi \cos \left(i\right) /D^2 $,
		 where $i$ is the inclination angle (defined as the angle between the 
		 observer's line-of-sight and the disk normal) and $D$ is the distance to the observer.
		 We can therefore write 
		\begin{align}
		w\left(E, t\right) =& \int_{0}^{2\pi} \int_{r_{\textmd{in}}}^{r_{\textmd{out}}}
		 K\left(r\right) g ^3\left(r,\phi\right)\delta\left(t-\tau\left(r,\phi\right)\right)\nonumber\\
		&\mathscr{R}\left(E/g(r,\phi)|\Gamma\right)  r \, {\rm d}r\,{\rm d}\phi, 
		\label{eq:impulse_response} 
		\end{align}
		where $K\left(r\right) \equiv \epsilon\left(r\right)\cos\left(i\right)/D^2  $. 
		A simple example of such a response function is reported in Fig.~\ref{fig:impulse_response}, 
		where $\mathscr{R}$  is a $\delta-$function at $6.4$ keV. 
		Although this is an over-simplification of 
		the restframe spectrum, it allows us to see the modifications to a narrow emission 
		line as a function of time and energy. 
		A more realistic scenario (Fig.~\ref{fig:impulse_response_xillver}) 
		is when $\mathscr{R}$ is calculated using \textsc{xillver}.
		The response function is drawn in the central panels of both 
		the two figures, while the sides panels represent the \textit{time averaged spectrum} 
		(right side panels) and \textit{impulse-response function} 
		i.e. the response function integrated over energy (bottom panels). 
		We will describe these plots in detail with all the parameters 
		used to compute $w\left(E, t\right)$ in Section~\ref{sec:simulation}.
		
		With the response function formalism we are able to write the time 
		dependence of the observed reflection spectrum 
		in terms of a convolution, which by the convolution 
		theorem corresponds to a multiplication in the Fourier domain. 
		The time dependence of the reflection spectral shape arises 
		entirely due to the response function which 
		provides the corrections to the restframe energy spectrum.
		The linearity assumption (i.e. $\beta=0$) has been used 
		for most previous reverberation mapping
		studies \citep[e.g.][]{Cackett2014,Emmanoulopoulos2014}. In the
		following section, we introduce for the first time a non-linear
		effect (i.e. $\beta \neq 0$).
	
		\subsection{Non-linear effects}
		\label{sub_sec:non-linear} 
		In the previous Section we set $\beta=0$ as has been done in the
		literature. This means that the response function has been calculated 
		assuming there are no phase lags associated with
		the continuum. 
		However,  neither AGN nor BHBs show any feature 
		in the lag spectrum indicative of reverberation lags
		for frequencies below $\sim 300 [M_\odot/M] \,{\rm Hz}$, with the lag spectrum 
		instead \textit{dominated} by featureless continuum lags 
		\citep[e.g.][]{Kotov2001,Walton2013}.
		
		We model non-linear variation of the continuum spectrum with a variation of the
		power-law index; i.e. $\beta \neq 0$. 
		We can see from Eq.~\ref{eq:refl_patch} that the shape of the reflection 
		spectrum now depends on time not just because of the response function,
		but also because it	depends on the variable power-law index. 
		Therefore Eq.~\ref{eq:refl_patch} becomes 
		\textit{non-linear} for $\beta \neq 0$ and the
		simple transfer function formalism is not useful any more. 
		We can, however, linearise.
		Following \citet{Kotov2001}, we can Taylor expand the
		continuum spectrum to get
		\begin{equation}
		D\left(E,t\right) \simeq A\left(t\right) E^{-\Gamma}
		{\rm e}^{-\nicefrac{E}{E_{\textmd{cut}}}} \left[1+\beta\left(t\right)\ln E
		\right],
		\label{eq:Dlinearization}
		\end{equation}
		where we keep terms up to first order. We can take this further, and
		also Taylor expand the restframe reflection spectrum
		\begin{equation}
		\mathscr{R}\left(\Gamma-\beta\left(t\right)|E\right) \simeq \mathscr{R}\left(E|\Gamma\right) - \beta\left(t\right) \frac{\partial\mathscr{R}\left(E|\Gamma\right)}{\partial\Gamma},
		\label{eq:Rlinearization} 
		\end{equation}
		where we compute $\partial\mathscr{R}/\partial\Gamma$ numerically.
		We explicitly test if it is reasonable to ignore higher
		order terms in the Taylor expansion in Appendix~\ref{app:linearization}.
		Since these expressions are both linear, we can
		once again employ the response function formalism using 
		Eq.~\ref{eq:Rlinearization} to obtain a second 
		response function (i.e. first order approximation whereas $w$ is the zero-order approximation): 
		\begin{align}
			w_1\left(E,t\right) = & \int_{0}^{2\pi} \int_{r_{\rm in}}^{r_{\rm out}} K\left(r\right) g^3\left(r,\phi\right)\delta\left(t-\tau\left(r,\phi\right)\right)\nonumber\\
			&\frac{\mathscr{R}\left(E/g(r,\phi)|\Gamma_2\right)- \mathscr{R}\left(E/g(r,\phi)|\Gamma_1\right) }{\Gamma_2-\Gamma_1}  r \, {\rm d}r\,{\rm d}\phi , 
		\label{eq:resp_func2}
		\end{align}
		where $ \Gamma_2=\Gamma+\nicefrac{\Delta\Gamma}{2}$, $
		\Gamma_1=\Gamma-\nicefrac{\Delta\Gamma}{2}$ and $\Delta\Gamma$ 
		sets the $\Gamma$ range where the variation of the restframe reflection spectrum as a 
		function of  $\Gamma$ is supposed to be linear. 
		To keep the linearity even after the Fourier transform we 
		define $B\left(t\right)=A\left(t\right) \beta\left(t\right)$. 
		Since $\beta$ is arbitrary, we are free to introduce the more useful arbitrary variable $B$. 
		The total emitted spectrum is simply $S(E,t)=D(E,t)+R(E,t)$ 
		and its Fourier transform after the linearization can be written as 
		\begin{align}	
			S\left(E,\nu\right) =& A\left(\nu\right)\left[ E^{-\Gamma} {\rm e}^{-\nicefrac{E}{E_{\textmd{cut}}}}+  W\left(E,\nu\right) \right]  \nonumber \\ 
			+ &B\left(\nu\right) \left[E^{-\Gamma} {\rm e}^{-\nicefrac{E}{E_{\textmd{cut}}}} \ln E - W_1\left(E,\nu\right) \right].
			\label{eq:Four_cont+refl}
		\end{align}
		Here, $W$ and $W_1$ are the \textit{transfer functions} i.e. Fourier
		transform of the response function. 
		Since the equation is linear in time the convolution 
		converts into a multiplication in the frequency domain. 
		We see the light-crossing lags are accounted for in Eq.~\ref{eq:Four_cont+refl} 
		by the transfer functions, and the phase
		difference between $A$ and $B$ at each frequency
		introduces a continuum lag (the explicit derivation of  
		Eq.~\ref{eq:Four_cont+refl} is in Appendix~\ref{app:linearization}). 
		
				We calculate the model complex covariance $ G(E,\nu) $ 
				by multiplying $ S(E,\nu) $ (Eq.~\ref{eq:Four_cont+refl}) with the complex conjugate of the 	
				Fourier transform $ F(\nu) $ of an arbitrary reference band 
				and dividing by its modulus. In the final expression of the complex covariance,  		
		we define the phase angles 
		$\phi_A(\nu)=\arg[A(\nu)F^*(\nu)]$ and $\phi_B(\nu)=\arg[B(\nu)F^*(\nu)]$. 
		We also define $\alpha(\nu)=\sqrt{\Delta} |A(\nu)|$ and 
		$\gamma\left(\nu\right)= |B\left(\nu\right)|/|A\left(\nu\right)|$. 
		The model complex covariance then becomes
		\begin{align}
		G\left(E,\nu\right)=\alpha\left(\nu\right)
		{\bigg [ }&{\rm e}^{i\phi_A\left(\nu\right)}
		\left[E^{-\Gamma} {\rm e}^{-E/E_{\rm cut}}+ W\left(E,\nu\right) \right]+\nonumber\\
		\gamma\left(\nu\right)\, &{\rm e}^{i\phi_B\left(\nu\right)}
		\left[E^{-\Gamma} {\rm e}^{-E/E_{\rm cut}} \ln E - W_1\left(E,\nu\right)\right]
		{\bigg ] }. 
		\label{eq:complex_covariance_model}
		\end{align}
		Therefore, for each Fourier frequency considered, we model the 
		continuum variation with four arbitrary parameters: 
		$\alpha\left(\nu\right)$, $\gamma\left(\nu\right)$, 
		$\phi_A\left(\nu\right)$ and $\phi_B\left(\nu\right)$.
				In the fits, these parameters are constrained by the data, 
				but due to their arbitrary nature they do not constrain any of the physical 
				parameters in our model. 
				This does not affect our conclusions on the reflection parameters, 
				which are constrained by the spectral correlations in the data described by our model.
		In principle, we could recover from these parameters the Fourier 
		transform of $\beta\left(t\right)$.
		However, in practice this involves a deconvolution ($B(\nu)=A(\nu) \otimes \beta(\nu)$), 
		which requires knowledge of the higher order variability properties of 
		$A(\nu)$ and $B(\nu)$ \citep{Kording2004}, which we do not constrain in our modelling. 
		Since this is simply a mathematical model however, $\beta(t)$ is of no more physical 
		interest than $\gamma(\nu)$, which can easily be constrained from data.
		To fit to the data Eq.~\ref{eq:complex_covariance_model} should be convolved 
		with the instrument response matrix converting energies $ E $ into channel numbers $ I $. 
		Therefore the final expression of the complex covariance model is 
		\begin{equation}
		G(I,\nu) = \alpha(\nu) \left[ {\rm e}^{i \phi_A(\nu)} Z(I,\nu) +
		\gamma(\nu) {\rm e}^{i \phi_B(\nu)} Z_1(I,\nu) \right],
		\label{eq:complex_covariance_channels}
		\end{equation}
		where $ Z(I,\nu) $ and $ Z_1(I,\nu) $ are respectively the convolution of 
		$Z(E,\nu)=E^{-\Gamma}{\rm e}^{-E/E_{\rm cut}} + W(E,\nu)$ and 
		$Z_1(E,\nu)=E^{-\Gamma}{\rm e}^{-E/E_{\rm cut}} \ln E - W_1(E,\nu)$
		with the instrument response (see Appendix~\ref{app:fit_data}  
		for a demonstration that convolving the complex covariance with the response 
		is equivalent to the actual process where we cross Fourier transforms of the convolved time series). 
		Note that any  additive/multiplicative model (such as line-of-sight 
		absorption) should be applied before the convolution operation,
		as shown in the previous section and in 
		Appendix~\ref{app:cross-spectrum_analysis}. 
		We implement  this by multiplying the real and imaginary part 
		of the complex covariance with the absorption component directly and 
		then convolving with the instrument response.

\section{Model Parameter Exploration}
\label{sec:simulation}
In this Section we explore the parameter space of our model. 
Although in the following Section we will use real and imaginary parts 
of the complex covariance to fit to data, here we consider time lags and 
variability amplitude to explore the parameter space, since these are more intuitive. 
We can consider two main groups of model parameters: those that govern the response function 
and therefore the reverberation lags, and those that govern the continuum lags. 
Further parameters govern the restframe reflection spectrum, which we only briefly discuss 
but refer the interested reader to \citet{Garcia2013}. 
Here, we first summarise the response function parameter dependencies 
and then concentrate on the continuum parameters. 
We then analyse the importance of accounting for the non-linear 
effects caused by fluctuations in the reflection energy spectrum.

		\subsection{Response function}
		\label{subsec:response_function}
		
		The response function for a lamppost geometry depends on the height $h$ 
		of the point source, the inclination angle $i$, 
		the inner ($r_{\rm in}$) and the outer ($r_{\rm out}$) radius of the disc, 
		the dimensionless spin parameter $a$ and the mass $M$ of the black hole. 
		The parameter dependencies of such a response function have already been extensively explored 
		in the literature \citep{Cackett2014,Emmanoulopoulos2014}. 
		Therefore, here we only briefly explore the transfer function and refer the interested 
		reader to previous papers for more detail. 
		In Fig.~\ref{fig:impulse_response}, we show the response function assuming that the 
		restframe reflection spectrum is simply a $\delta-$function iron line at $6.4$ keV. 
		In Fig.~\ref{fig:impulse_response_xillver} we instead use \textsc{xillver} to calculate the restframe reflection spectrum, 
		setting the iron abundance $A_{\rm Fe}=1$, ionisation parameter $\log\xi =3.1$, cut-off energy of the incident 
		power-law $ {\rm E}_{\rm cut} = 300$ keV and reflection fraction to $1.0$.  
		
		For both figures, we set $i=45^\circ$, $h=10$, $r_{\rm out}=10^6$ and $a=0.998$. 
		Since we represent time in units of $R_{\rm g}/c$, Fig.~\ref{fig:impulse_response} 
		and Fig.~\ref{fig:impulse_response_xillver} are independent of black hole mass. 
		In Fig.~\ref{fig:impulse_response} blue, green and orange represent an inner radius of $r_{\rm in}=1.2$, $10$ and $50$ 
		respectively, while Fig.~\ref{fig:impulse_response_xillver} shows the  $r_{\rm in}=10$ case only. 
		The central panels show the response function (with shades representing flux), 
		the bottom panels show the impulse response function, 
		and the right hand panel shows the time-averaged line profile. 
		The time axis is defined such that the $\delta-$function flash in the 
		continuum reaches the observer at a time of zero $R_{\rm g}/c$.
		After the continuum flash, the next photons to reach the observer are those that reflect from the front of the 
		disc (as seen by the observer), at a radius that depends on $h$ and $i$. 
		For $i = 45^{\circ}$ and $h= 10 $ this radius is $10 R_{\rm g}$. 
		The initial sharp rise in the blue and green curves in Fig.~\ref{fig:impulse_response} 
		therefore occurs at the same time because the inner disc radius for both cases is 
		$\leqslant 10 R_{\rm g}$.
		The secondary peak in the impulse response function indicates 
		when we see the first photons that reflect from the back of the disc.
		For a small $r_{\rm in}$, the broadest iron line is seen shortly after the continuum flash 
		($\sim 15 R_{\rm g}/c$ for $r_{\rm in}=1.2$), whereas for $r_{\rm in}=50$ the iron 
		line is initially narrow with its width peaking at $\sim 60 R_{\rm g}/c$. 
		This is because gravitational redshift is important for small radii, whereas 
		Doppler broadening is dominant for larger radii.
		This can also be seen in the time-averaged line profile, 
		which is smeared and skewed when the inner radius is small (e.g. blue line 
		$r_{\rm in}$ = 1.2), while it has the characteristic double horn profile primarily due to 
		Doppler shifts when the disc is far from the black hole (e.g. orange-red line $r_{\rm in} = 50$ ). 
		If we compare Fig.~\ref{fig:impulse_response} with Fig.~\ref{fig:impulse_response_xillver}, 
		we see the response functions and line profiles are very different because we used 
		a different restframe reflection spectrum (we must compare the green lines because they have the same radius). 
		In contrast, the restframe reflection spectrum makes little difference to the impulse response function, 
		since this is the integral of the response function over all energies.

		\subsection{Continuum Variability}
		\label{(subsec:continuum_variability)} 
		
		The continuum emission depends on the power-law index $\Gamma$ and cut-off energy $E_{\rm cut}$. 
		We see from Eq.~\ref{eq:complex_covariance_model} that the continuum variations
		in a frequency range $\nu-\Delta/2$ to $\nu+\Delta/2$ are governed by the parameters 
		$\alpha\left(\nu\right)$, $\gamma\left(\nu\right)$, $\phi_A\left(\nu\right)$ and $\phi_B\left(\nu\right)$. 
		$\alpha(\nu)$ is simply a normalization of the variability amplitude for each frequency; 
		i.e. it does not affect the phase lags at all and does not affect 
		the energy dependence of the variability amplitude. 
		Together, the other three parameters govern the energy 
		and frequency dependence of the phase lags. 
		We note that the four continuum variability parameters are not independent 
		and one of them can in principle be derived from the other three using 
		the definition of the reference band 		
		(see Appendix~\ref{app:model_parameter}). 
		For the illustrative examples presented in this Section, we always use $\phi_A\left(\nu\right)=0$. 
		This would, for example, be the case if there were no contributions from reflection 
		and the reference band were chosen to be at $1$~keV.		
 
\begin{figure*}
	\includegraphics[width=\textwidth]{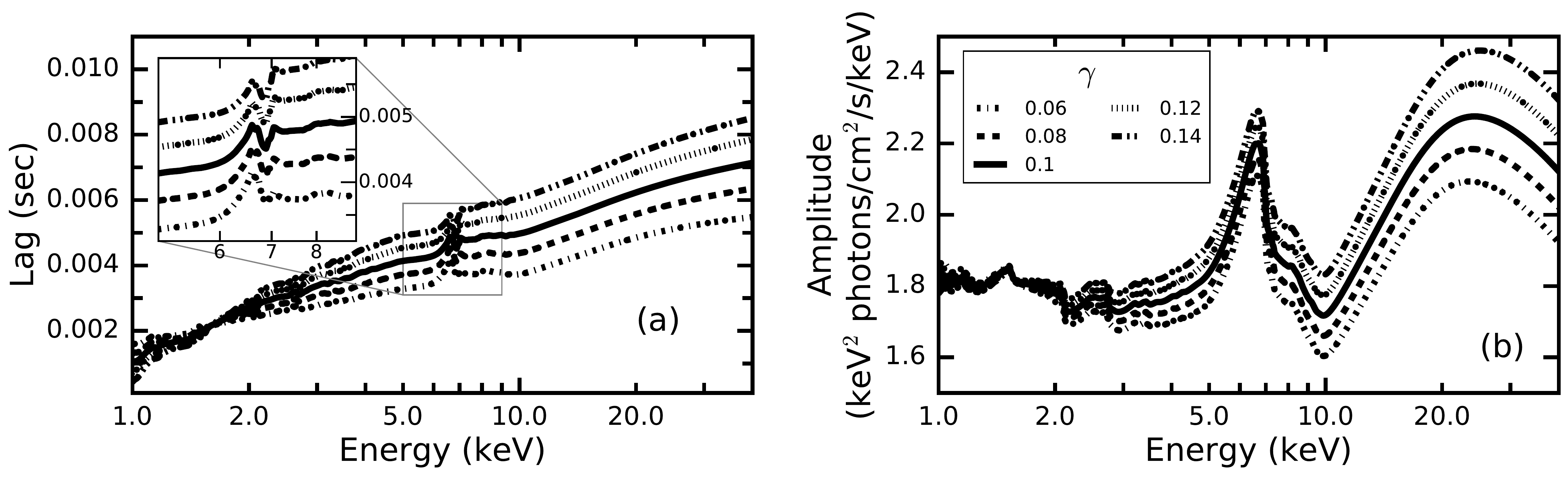}
	\caption{	 
		 Predicted time lags (a) and absolute variability amplitude (b) as a function of energy for the frequency range $1-2\, {\rm Hz}$. 
		 For both panels, $\phi_A=0\,{\rm rad}$ and $\phi_B = 0.2 \,{\rm rad}$. The curves have different values of $\gamma$: $0.02$ solid line, $0.04$ dashed line, $0.06$ dotted dashed line, $0.08$ dotted line, $0.1$ dashed double dotted line.
		 In both panels we use the parameters: $\Gamma = 2$, $i=45\degree$, $r_{\textmd{in}}=10$, $r_{\textmd{out}}=10^6$,
		  $h=10\,R_{\rm g}$, $a=0.998$, $M=10M_{\odot}$, $\log_{10}\xi = 3.1$, $A_{\rm Fe}=1$ 
		(respectively ionization and iron abundance in \textsc{xillver}).
		}
	\label{fig:lag_amp_g}
\end{figure*}

\begin{figure*}
	\includegraphics[width=\textwidth]{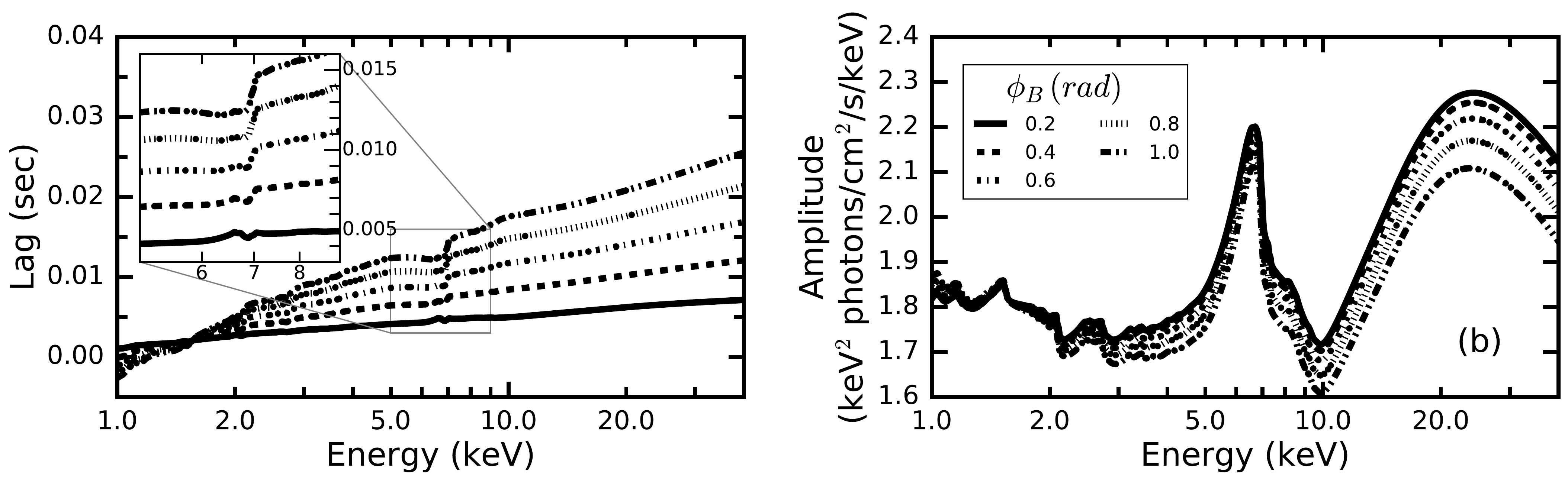}
	\caption{
		Predicted time lags (a) and absolute variability amplitude (b) as a function of energy for the frequency range $1-2\, {\rm Hz}$. 
		For both panels, $\phi_A=0\, {\rm rad}$ and $\gamma = 0.1$. 
		The curves have different values of $\phi_B$: $0.2\,{\rm rad}$ solid line, $0.4\,{\rm rad}$ dashed line, $0.6\,{\rm rad}$ 
		dotted dashed line, $0.8\,{ \rm rad}$ dotted line, $1 \,{\rm rad}$ dashed double dotted line. 
	 	 In both panels we use the parameters: $\Gamma = 2$, $i=45\degree$, $r_{\textmd{in}}=10$, $r_{\textmd{out}}=10^6$,
		 $h=10\,R_{\rm g}$, $a=0.998$, $M=10M_{\odot}$, $log_{10}\xi = 3.1$, $A_{\rm Fe}=1$ 
		 (respectively ionization and iron abundance in \textsc{xillver}).
		}
	\label{fig:lag_amp_phiB}
\end{figure*}

		To explore the continuum parameters, we set the black hole mass to $10 M_\odot$ 
		and fix the reflection parameters to those used for Fig.~\ref{fig:impulse_response_xillver}. 
		In Figs.~\ref{fig:lag_amp_g} and \ref{fig:lag_amp_phiB}, we show the time lag (a) and 
		variability amplitude (b) as a function of energy for a frequency range $1-2$ Hz. 
		Here, amplitude is \textit{absolute} amplitude in units of energy flux, not fractional amplitude. 
		In Fig.~\ref{fig:lag_amp_g}, we fix $\phi_B\left(\nu\right)=0.2$ rad and different lines correspond 
		to different values of $\gamma\left(\nu\right)$, which controls the amplitude of the power-law index oscillation. 
		We see that increasing $\gamma\left(\nu\right)$ gives a larger absolute value of the lag. 
		We can see from Eq.~\ref{eq:complex_covariance_model} that setting $\gamma\left(\nu\right)=0$ 
		leads to no continuum lag at all. 
		Fig.~\ref{fig:lag_amp_g}b shows that the amplitude spectrum becomes harder when $\gamma$ is increased. 
		We can partially understand this by imagining a power-law pivoting around some energy $E_0$, 
		such that the flux at $E_0$ remains constant and 
		the variations for $E>E_0$ are in anti-phase with the variations for $E<E_0$. 
		In this case, the variability amplitude increases with $|E-E_0|$. 
		For a more realistic case in which the power-law index \textit{and} the normalisation are varying 
		(with a general phase difference between the variations in power-law index and normalisation), 
		there can be no energy at which there is \textit{zero} variability amplitude, 
		but there will be an energy at which the amplitude reaches a \textit{minimum}. 
		The phase lags will also be anti-symmetric about this energy. 
		We will call this the pivot energy, $E_0(\nu)$. 
		Note that the pivot energy can be a function of frequency. 
		Since the amplitude increases with energy in Fig.~\ref{fig:lag_amp_g}, 
		the pivot energy for this example is $E_0(\nu)< 1$~keV.  
		It is worth noting however that extra complication occurs when a realistic spectral 
		model including photoelectric absorption and a low energy cut-off are considered.
		
		In Fig~\ref{fig:lag_amp_phiB}, we fix $\gamma(\nu)=0.1$ and vary $\phi_B(\nu)$. 
		We see that increasing $\phi_B(\nu)$ also increases the lags. 
		Panel (b) shows that increasing $\phi_B(\nu)$ makes the amplitude spectrum softer. 
		This can be understood partially as the pivot energy increasing as $\phi_B(\nu)$ is increased. 
		Figs. \ref{fig:lag_amp_g}a and \ref{fig:lag_amp_phiB}a show features 
		around the iron line in the lag-energy spectrum, which are highlighted with a zoom-in. 
		These result from the combination of reverberation lags and continuum lags, and also non-linear effects. 
		The simplest effect comes from the reverberation lag simply being a different value from the continuum lag. 
		In this case, the total energy spectrum is continuum plus reflection and so 
		we expect a dip in the lags at the iron line if the reverberation lag is smaller than the continuum lag,
		and a peak for the opposite case, since the reflection dominates the energy spectrum 
		at the iron line energy band. 
		Non-linear effects further contribute to these features, with the variations in continuum power-law index
		causing changes in the shape of the disc reflection energy spectrum. 
		This combination of effects leads to the dependence of the lag around 
		the iron line on $\gamma(\nu)$ and $\phi_B(\nu)$ being subtle.

		\subsection{The Importance of Non-Linear Effects}
\begin{figure*}
	\includegraphics[width=\textwidth]{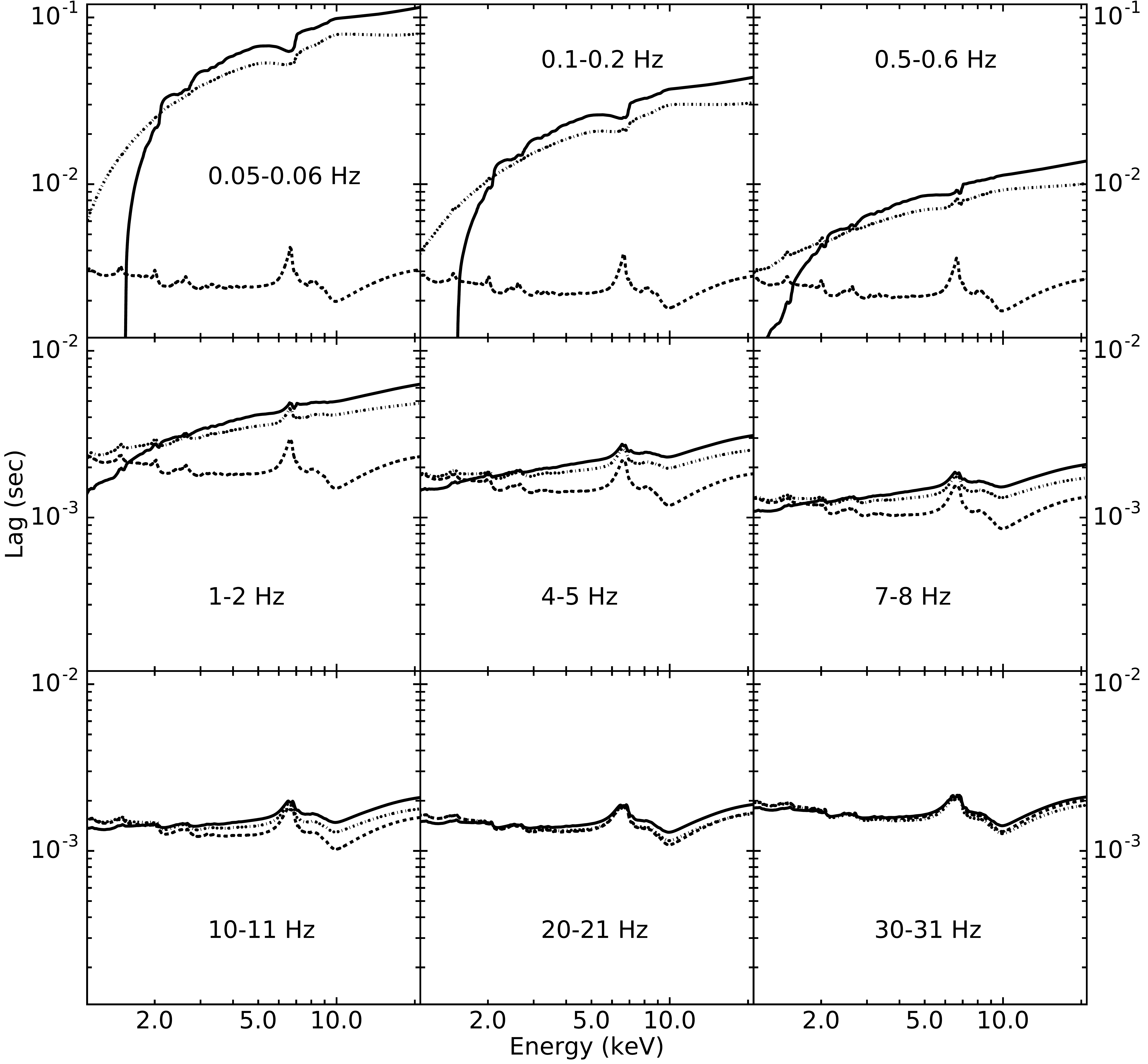}
	\caption{Time lag as a function of energy for different frequency ranges. 
		The solid lines represent the lag calculated with the complete model: 
		pivoting power-law in the continuum emission and contribution of reverberation. 
		The dashed line is the lag only due to the reverberation 
		(no pivoting power-law in the continuum emission). The dash-dotted line is the lag
		calculated considering the non-linear effect in the continuum emission (pivoting power-law)
		but naively ignoring this effect in the reverberation lag. 
		The other parameters are $\phi_A=0\,{\rm rad}$, $\phi_B=0.2\,{\rm rad}$, 
		$\Gamma = 2$, $i=45\degree$, $r_{\textmd{in}}=10$, $r_{\textmd{out}}=10^6$,
		$h=10$, $a=0.998$, $M=10M_{\odot}$, $\log_{10}\xi = 3.1$, $A_{\rm Fe}=1$.
		}
	\label{fig:lag_fq}
\end{figure*}

		Figure~\ref{fig:lag_fq} shows the lag spectrum for different frequency ranges. 
		The solid lines are produced with the complete model (both continuum and reverberation lags), 
		while the dashed lines only include the reverberation lag 
		(there is no variation of the continuum power-law index). 
		For the dashed dotted lines, we instead calculate the continuum lags
		using our pivoting power-law model, but naively do not account for 
		the variations in continuum power-law index when calculating 
		the reflection spectrum (i.e. we set $W_1 = 0 $ artificially). 
		This allows us to assess the bias caused by not self-consistently accounting 
		for the non-linear nature of the continuum variations. 
		Here, we fix $\gamma(\nu)$, $\phi_A(\nu)$ and $\phi_B(\nu)$ to be constant 
		with frequency (see the figure caption for all parameter values) and assume 
		the same reflection parameters as those used for the previous sub-section. 
		We see that, consistent with observational data, the continuum lags dominate 
		over the reverberation lags for low frequencies, 
		and the decrease of the continuum lag with frequency allows 
		the reverberation lag to dominate at the highest frequencies. 
		For low frequencies, the full model does differ quite significantly from the naive treatment, 
		particularly below $\sim 2$ keV for which the full model predicts 
		a steep break in the lag-energy spectrum. 
		Therefore, ignoring non-linear effects could bias measurements of the reverberation lags. 
		At high frequencies the models converge, since the lags are dominated by reverberation lags.
		
\begin{figure}
	\includegraphics[width=\columnwidth]{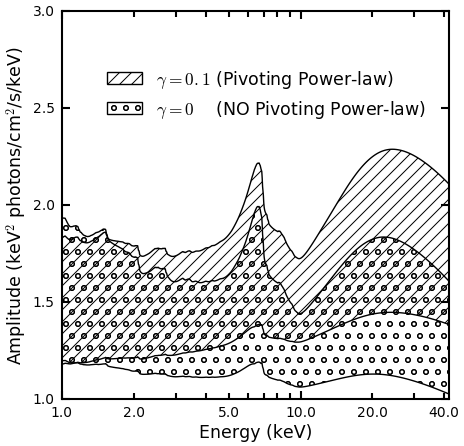}
	\caption{Covariance amplitude as a function of energy. 
		The two regions represent frequency ranges from 
		$0.02\, {\rm Hz}$ (the two highest curves) to $50 \, {\rm Hz}$ (the two lowest curves). 
		For the hatched region, we use the full model (pivoting power-law) 
		and for the dotted region we do not include any continuum lags (NO pivoting model).
		All the parameters are equal to the ones in Fig.~\ref{fig:lag_fq}.
		}
	\label{fig:amp_fq}
\end{figure}

		Fig.~\ref{fig:amp_fq} shows the modulus of the complex covariance for a range of Fourier 
		frequencies for the linear model (i.e. $\gamma= 0$, meaning there are no continuum lags) 
		and the full model ($\gamma= 0.1$, meaning there are now continuum lags), 
		represented respectively by the dots and the hatching. 
		We see that, for both cases, the iron line feature is stronger 
		for low frequencies (top lines) than for high frequencies (bottom lines). 
		This is a result of the finite size of the reflector. 
		Whereas the continuum can vary on arbitrarily short timescales, 
		the fastest variability is washed out in the reflected emission by path length differences 
		between rays that reflected from different parts of the disc \citep[][]{Gilfanov2000,Cackett2014}.
		We also see that in terms of amplitude there is little difference between the two models, 
		aside from the higher overall variability amplitude for the full model 
		introduced by fluctuations in the continuum power-law index. 
		The non-linear effects considered here therefore influence the predicted lags more than the amplitude
		and could therefore easily be missed when ignoring the lags.

\section{Example fits to Cygnus X-1 data}
\label{sec:fit_data}

As a proof of principle of our method, we fit the complex covariance for a few hard state 
observations of Cygnus X-1 for multiple Fourier frequencies. 
		The data analysis presented here is intended primarily to show the feasibility 
		of jointly describing the non-linear continuum and resulting reflection 
		variability using the methods outlined above. 
		As the model does not contain a description of the relativistic light bending, 
		the fit results should be interpreted with caution,  
		in particular if they require a corona/disc geometry close to the black hole. 
Cygnus X-1 is one of the first black hole X-ray binaries observed and has been studied 
extensively during the past decades, in particular with \textit{the Rossi X-ray Timing Explorer} (\textit{RXTE}). 
Another motivation to select this source is the absence of strong low frequency QPOs, 
which are routinely seen in many other black hole X-ray binaries. 
Since there is good evidence that these signals are due to precession of the inner accretion 
flow \citep{Ingram2016}, or at least geometrical in origin 
\cite[e.g.][]{Heil2015,Motta2015,Eijnden2017}, they complicate the situation somewhat. 
In particular, transfer modelling assumes a stationary geometry, 
and so it is convenient to avoid observations with strong QPOs. 
Following \citet{Revnivtsev1999,Kotov2001}, we use \textit{RXTE} observations P10238 
recorded between March $26^{\rm th}$ and $31^{\rm st}$ 1996 by the proportional counter array (PCA). 
In all, this consists of seven observations\footnote{
					Observation IDs: 10238-01-05-00, 10238-01-05-000, 10238-01-06-00, 10238-01-07-00, 10238-01-07-000, 10238-01-08-00, 10238-01-08-000.}.
					However, since the two first observations\footnote{Observation IDs: 10238-01-05-00,10238-01-05-000.}  
					have a slightly different spectral slope we only consider these last five.		
						
The timing data are in the `Generic Binned' mode $B\_16{\rm ms}\_64{\rm M}\_0\_249$, 
which has $1/64$ sec time resolution in $64$ energy channels covering the whole PCA energy band. 
We apply standard \textit{RXTE} good time selections (elevation greater than $10^\circ$ and offset less than $0.02^\circ$) 
and additionally select times when $5$ proportional counter units (PCUs) were switched on, 
using \textsc{ftools} from the \textsc{heasoft} 6.19 package. 
This gives a total exposure time of $56.2$ 	
ks (after sorting into segments of contiguous data, 
	the remaining useful exposure is $ 46.6 $ ks).
Following the procedure outlined in Section 2, we calculate the complex covariance for $8$ 
frequency ranges between $0.017$ and $32$ Hz.
We assume unity coherence for this observation, which is a good assumption for the hard state 
 of Cygnus X-1 \citep{Nowak1999,Grinberg2014}.
The reference band is always $2.84 - 3.74$ keV and 
we consider the energy range $4 - 25$ keV for fitting.
	The choice of the reference band is discussed in Appendix~\ref{app:model_parameter}.

	We use \textsc{xspec} version 12.9  \citep[][]{Arnaud1996} to fit real and imaginary parts 
	of the complex covariance as a function of energy, 
	simultaneously for all $8$ frequency ranges considered.
	We created an \textsc{xspec} local model for the complex covariance 
		following the procedure explained in Section~\ref{sec:model} 
		using a cut off power-law for the continuum spectrum and
		\textsc{xillver} for the rest-frame reflection spectrum.	
		We additionally fit the time-averaged energy spectrum with the 
		DC component of the covariance model,
		meaning that we simultaneously fit accross $17$ spectra. 
		The energy spectrum is computed adding the PCA standard $2$ data of the considered 
		observations with the ftool \textit{addspec} and adding in quadrature 0.1\% systematic error
		\footnote{We add systematic errors only in the time-average energy spectrum. 
				The complex covariance spectra are 
				dominated by statistical errors so that no systematic errors are needed.}.
	 
	For each of these 17 spectra, absorption is accounted for 
	using the multiplicative model \textsc{TBabs},
	assuming the abundances of \citet{Wilms2000}.
	All parameters are tied to be the same for real and imaginary parts of a given frequency range. 
	The parameters governing the continuum variations, $\phi_A(\nu)$, $\gamma(\nu)$ 
	and $\phi_B(\nu)$, are allowed to vary freely with frequency. 
	For the time-averaged spectrum, $\phi_A=\phi_B=\gamma=0$, 
	whereas $\alpha(\nu)$ is a free normalisation parameter. 
	All remaining model parameters are tied to be the same for all $17$ spectra. 
	We fix the mass of the black hole to $14.8~M_\odot$ \citep[][]{Orosz2011}. 
	It is important to note that this analysis is sensitive to black hole mass 
	and can therefore in future be used as a new method to estimate the mass. 
	However, that is beyond the scope of this current paper.

	We achieved a reduced $\chi^2=480.64/460$ with the physical 
		parameters in Table~\ref{tab:disc_par}. 
		For the best fitting model, $ r_{\rm in}$ 
		is pegged at its lowest allowed value ($1.5$). 
		This means that the model is not able to fully reproduce the shape 
		and normalisation of the iron line simultaneously 
		over the full range of frequencies considered.
		The source height is very  
				close to the black hole ($ 2.4 \pm 0.5 $). 
				As noted above, the values of these two parameters 
				suggest caution in interpreting our results, 
				as we explore a region close to the black hole without 
				accounting for all the relevant physics (in particular, light bending). 
			 
		To simplify the explored parameter space and to mimic 
		an oblate geometry of the illuminating corona, 
		we additionally tried tying the inner radius 
		of the disc to be twice the height of the point source. 
		In this case, the best fit has a reduced $\chi^2=484.43/461$,  
		the height of the source is pegged to the lowest value. 
		An F-test favours the first of the two models.
		The hydrogen column density is a free parameter in our best fitting model, 
		with the best fit value giving a significantly better $ \chi^2 $ 
		than fixing $ N_{\rm h}=0.6\times 10^{22} {\rm cm}^{-2} $ following \citet{Gilfanov2000}.
		From Table~\ref{tab:disc_par} 
		it could be noted that our fit requires a strongly super-solar iron abundance $A_{\rm Fe} = 4.0\pm0.1$ 
		as has previously been found in this source \citep[e.g.][]{Duro2016}
		\footnote{Analyzing Cygnus X-1 soft state \citet{Tomsick2014} 
			also found evidence of super-solar abundance 
			although still much lower than our result.}  
			and other BHBs \citep[e.g. GX 339-4:][]{Garcia2015}
		\footnote{Although the empirical evidence for these super-solar iron abundances
			is now very strong, the physical cause is still unknown.
			It could result from radiative levitation of the iron atoms in the 
			inner disc, or perhaps is merely an artifact of some missing physics 
			in the current state-of-the-art reflection models.}.
		The high energy cut-off in our fit is compatible with what 
		\citet{Wilms2006} found for the hard state of Cygnus X-1. 
		We notice our value is slightly higher than their average, 
		but the authors used a different model to fit the spectrum, 
		for example accounting for the reflection with a Gaussian curve. 	
		We find a high reflection fraction compared to  e.g. \citet{Parker2015,Basak2017}.
		We expect this result to be highly biased by the absence of light bending in our model.

\begin{figure}
	\includegraphics[width=\columnwidth]{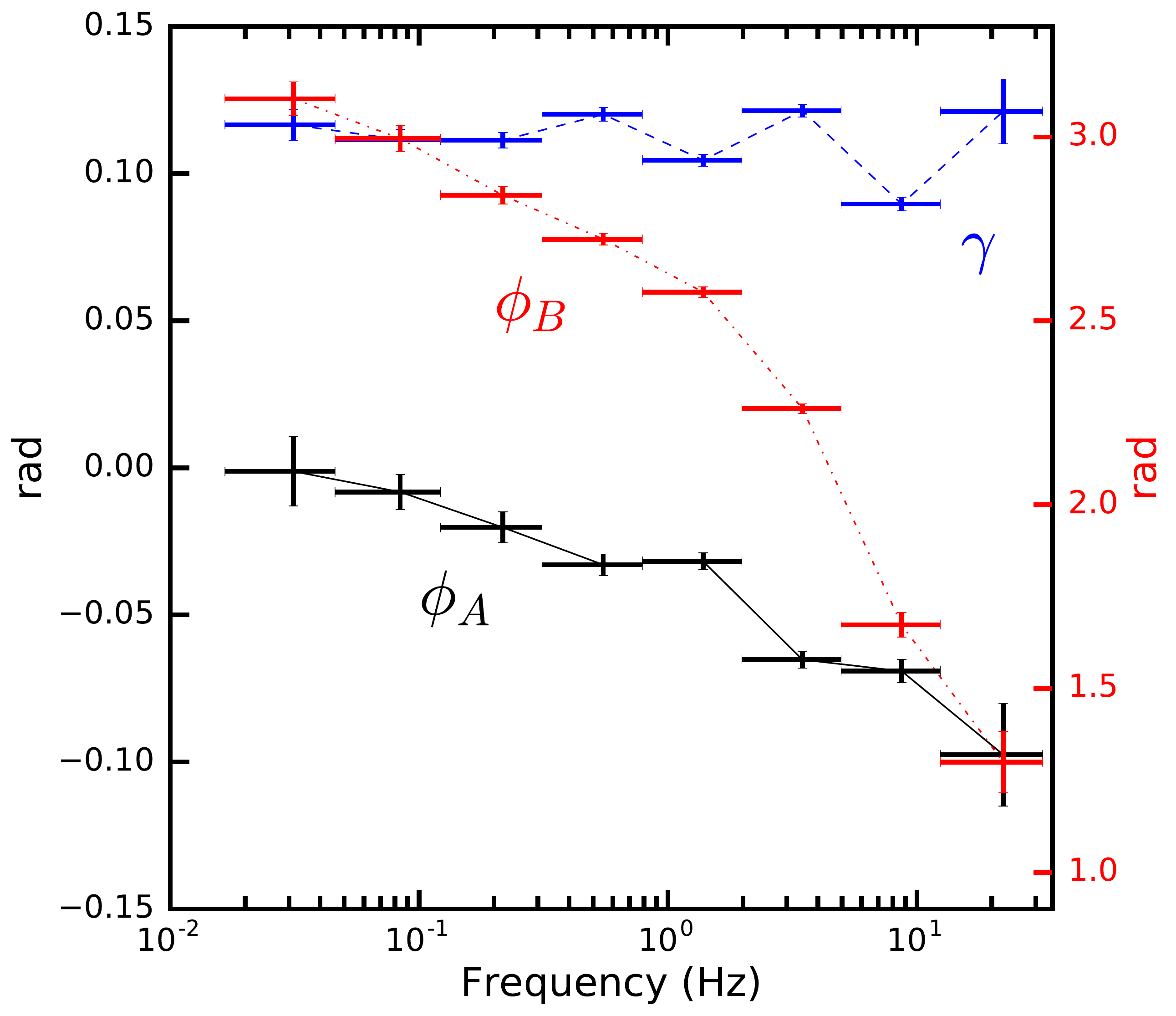}
	\caption{Continuum variability parameters as function of Fourier frequency. 
		The black solid and blue dashed curves refer to the left y-axis, while the red dotted curve 
		refers to the right y-axis. The x-errorbars represent the range of frequency used to fit 
		the complex covariance as a function of energy. The points with the 
		same frequency range are computed from the same fit.}
	\label{fig:pl_par}
\end{figure}

The best fitting continuum parameters are plotted in Fig.~\ref{fig:pl_par}. 
Here, the black and blue points ($\phi_A$ and $\gamma$ respectively) 
correspond to y-axis scale on the left, while the red points ($\phi_B$) 
correspond to the y-axis scale on the right. 
$\phi_A$ and $\phi_B$ are in units of radians, whereas $\gamma$ is dimensionless. 
We see that all $3$ parameters reduce with frequency. 
Since our continuum lag model is simply empirical, the direct 
physical meaning of these parameters is not immediately clear. 
It is still interesting to compare the results in Fig.~\ref{fig:pl_par} 
with more physical models.  

\begin{figure*}
	\includegraphics[width=\textwidth]{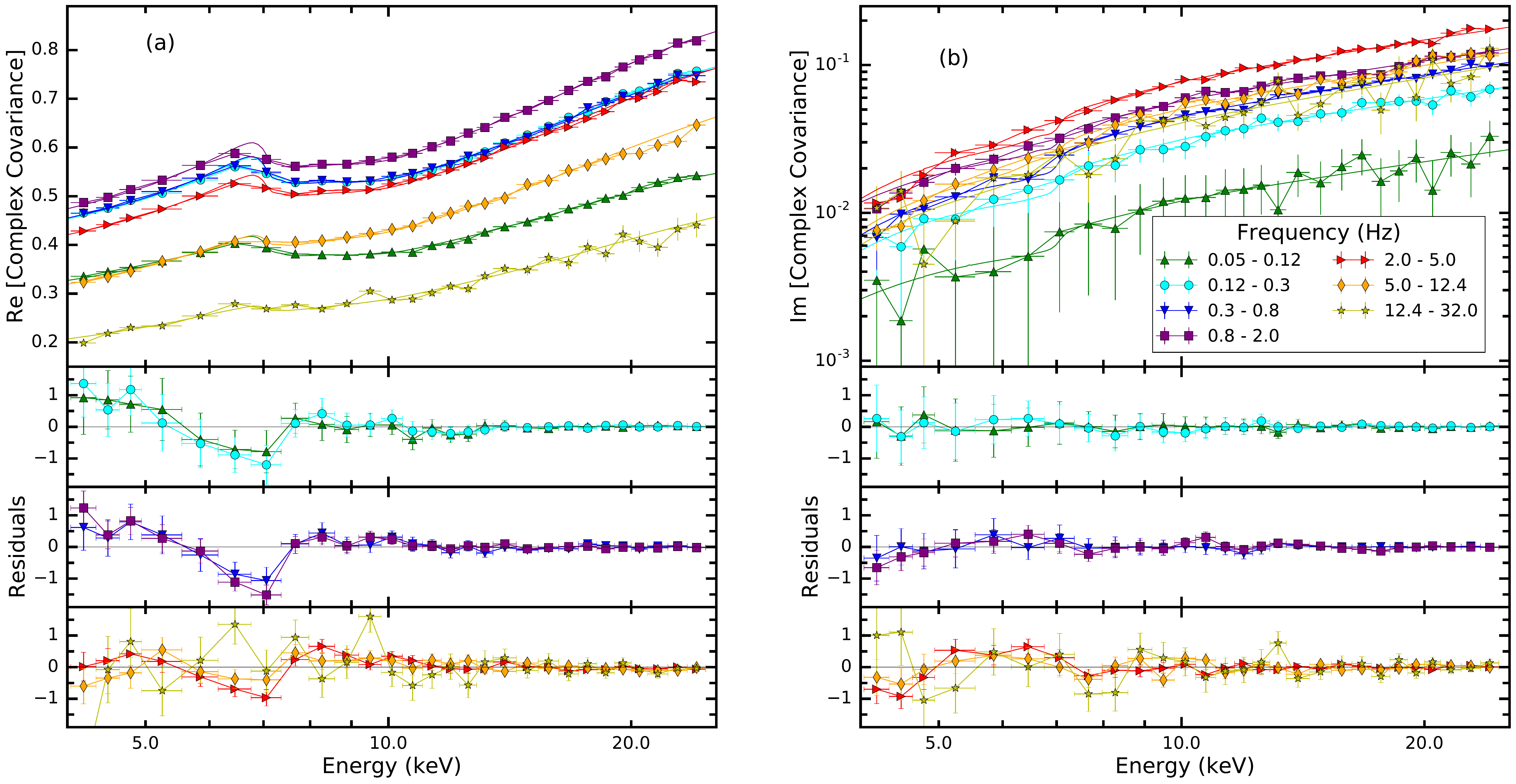}
	\caption{
		Fit of the real (a) and imaginary (b) part of Cygnus X-1 complex covariance 
		spectrum for different Fourier frequency ranges. 
		The dots are the data, the solid line is the model with a better energy resolution. 
		Both the bottom panels show the data minus the folded model 
		in units of normalised counts per second per keV (command residuals in xspec). 
		The residuals around the iron line are discussed in the text.
		}
	\label{fig:Cyg_re_im}
\end{figure*}

\begin{figure*}
	\includegraphics[width=\textwidth]{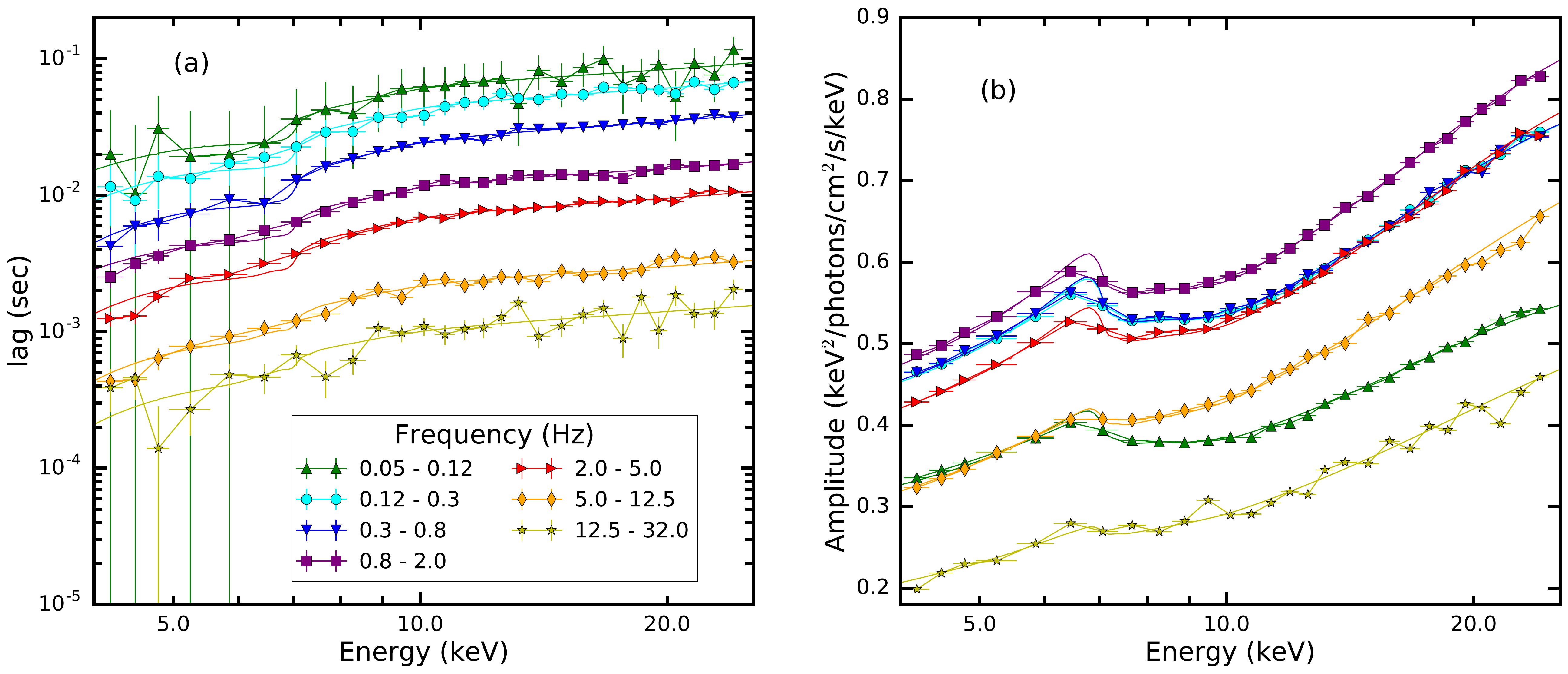}
	\caption{
		Lag (a) and variability amplitude (b) as a function of energy 
		for different Fourier frequency ranges. 
		Both data and model have been derived from the fit of real and imaginary 
		part of the complex covariance spectrum (Fig.~\ref{fig:Cyg_re_im}).
		}
	\label{fig:Cyg_lag_amp}
\end{figure*}

Figs.~\ref{fig:Cyg_re_im} and \ref{fig:Cyg_lag_amp} show 
the data and best fitting model for $7$ of the $8$ frequency 
ranges considered (the lowest frequency range is very noisy). 
In Fig.~\ref{fig:Cyg_re_im}, we show real (a) and imaginary (b) parts of the complex covariance, 
and plot the model with a higher energy resolution than the data for clarity. 
We additionally show fit residuals in the bottom panels. 
In Fig.\ref{fig:Cyg_lag_amp}, we instead represent the data 
and model as time lag (a) and variability amplitude (b). 
In the lags, we see the characteristic dip at $\sim 6.4$ keV for both 
data and model, which becomes less prominent for higher frequencies. 
This occurs mainly because the continuum lags are greater 
than the reverberation lags, and so at the iron line the greater contribution 
from reflection dilutes the overall time lag. 
For higher frequencies, the difference between continuum and reverberation 
lags reduces and so the dip becomes less prominent. 
For even higher frequencies, we expect to see an emission-like feature 
at the iron line, but unfortunately the maximum possible Nyquist frequency 
for this data mode is $32$ Hz. 
For the amplitude, we see the iron line becomes less prominent for higher frequencies, 
which is due to the finite size of the reflector as discussed in Section~\ref{sec:simulation}.

Fig.~\ref{fig:Cyg_re_im} shows that there are some systematic residuals in the 
real part of the complex covariance around the iron line. 
This seems to be because the amplitude of the iron line reduces more steeply 
with frequency in the data compared with the model. 
In the model, the accretion geometry (i.e. the source height and disc inner radius) 
sets both the frequency dependence of the iron line amplitude \textit{and} the width of the iron line. 
For a smaller inner disc radius, the iron line is broader and the variability 
amplitude of reflection drops-off less steeply with frequency. 
Therefore, it appears that our small best-fitting inner radius of $r_{\rm in} \approx 1.5$ 
reproduces the broad iron line seen in the complex covariance and time-averaged spectrum, 
but predicts a slower drop-off in reflection variability amplitude with frequency than is observed. 
Since the data constrain the broadness of the iron line at all frequencies better than they constrain
the drop off of the line amplitude at high frequencies, our fit returns a small value for inner radius. 
These residuals may be fixed in future by considering light bending, 
since this would give $1$) longer reverberation lags for a given 
disc inner radius due to longer path lengths of rays close to the black hole, 
and $2$) a broader iron line for a given disc inner radius due to the steeper 
emissivity profile resulting from focusing of rays close to the black hole.  

\begin{table*}
	\centering	
	\caption{Best fitting disc parameters obtained from our simultaneous fit to the complex 
		covariance in $8$ frequency ranges ($0.017 - 32 Hz$) and the time-average spectrum. 
		The $\chi^2$ is $ 480.64 $ with $ 460 $ degrees of freedom. 
		We report 1 $ \sigma $ errors for each parameter value.
		}
	\label{tab:disc_par}
	\begin{tabular}{  c c c c c c c c c } 
		\hline
		$ N_{\rm h} $ $\left(10^{22} \,\,{\rm cm}^{-2}\right)$ & $\Gamma$ & h $ (R_{\rm g})  $ &$ r_{\rm in} \, (R_{\rm g}) ^{a}$ & Incl $({\rm deg})$ & $A_{\rm Fe}$ & $E_{\rm cut}$ (keV)& $\log \xi $& Reflection Fraction$ ^b $\\
		\hline
		$0.2 \pm^{0.2}_{0.1}$  & $1.603 \pm^{0.004}_{0.003}$& $ 2.4 \pm^{0.5}_{0.5} $ & $ 1.5 \pm^{0.6} _{0} $& $35.7 \pm^{1.2}_{1.3}$	&$4.0 \pm^{0.1}_{0.1}$&$241 \pm^{10}_{5}$&$ 3.08 \pm^{0.01}_{0.01}$ &$- 0.93 \pm^{0.02}_{0.02}$\\		
		\hline
	\end{tabular}
	\begin{list}{}{}
		\item[$^a$] The parameter is pegged at its minimum allowed value.
		\item[$^b$] In \textsc{xillver} a negative reflection fractions means the model represents only the reflection spectrum without the continuum.
	\end{list}
\end{table*}

\section{Discussion}	
\label{sec:discussion}
		In this paper we have introduced a formalism to fully utilize X-ray reverberation 
		mapping as a tool to measure the geometry of accreting black holes, 
		and ultimately to provide a new means of measuring black hole mass and 
		inner radius of the disc.
		The main innovation is that we fit a reverberation model that considers 
		phase lags and variability amplitude jointly 
		for a wide range of Fourier frequencies for the first time. 
		In order to do this, we introduce the complex covariance, and show that 
		it is statistically more convenient to fit for the real and imaginary parts 
		of the complex covariance than fitting directly for the amplitude and 
		phase lags, which has led to some previous inaccuracies in the literature
		(it is still possible to compute the amplitude and phase lags, 
		which are more physically intuitive, and 
		compare them with other works).
		In order to fit for the full range of Fourier frequencies 
		(determined by the duration and time resolution of the observation)
		we need to account for the continuum lags that dominate at low frequencies. 
		We introduce a simple variation in the slope of the continuum emission 
		(pivoting power-law model), and use a Taylor expansion to calculate the model analytically whilst 
		still accounting for the variations in disc reflection spectrum shape caused by fluctuations 
		in the illuminating power-law index. 
		As noted by \citet{Kotov2001,Kording2004}, a pivoting power-law model is fairly attractive, 
		since it naturally produces time lags with a log-linear energy dependence, 
		as is regularly observed. 

		The continuum lags are often assumed to result from propagating mass 
		accretion rate fluctuations, with perturbations far from the black hole 
		propagating inwards and modulating new fluctuations generated 
		closer to the black hole \citep{Lyubarskii1997,Kotov2001,Arevalo2006}. 
		We note that, if the power-law emitting region is small compared with the disc inner radius, 
		our adopted lamppost geometry still provides a good approximation to propagating fluctuations models. 
		However, our treatment is at odds with sandwich models that consider propagation 
		in a coronal layer above and below the disc. 
		It could be that propagating fluctuations \textit{cause} fluctuations in power-law index, 
		through, for example, accretion rate fluctuations in the disc giving rise 
		to fluctuations in the luminosity of seed photons \citep[][Uttley \& Malzac in prep]{Uttley2014}. 
		The fluctuations in power-law index could also result from 
		fluctuations in electron temperature, seed photon luminosity and/or optical 
		depth of the Comptonising region.
		
		A few authors have made progress on integrating a
		propagating fluctuation model with reverberation modelling.
		\citet{Wilkins2013} consider fluctuations propagating through the corona with
		different geometries, and calculate reflection from each region of the corona. 
		This is perhaps the most physically self-consistent treatment in the literature, 
		but requires many response functions and so is computationally expensive. 
		It will therefore be challenging to attempt the multi-frequency fits that we present here with such a model.
		\citet{Chainakun2016a} and \citet{Chainakun2017} instead explore a `two blobs' model, 
		consisting of two lamppost sources with different heights and different intrinsic spectra, 
		with the propagation time between the two sources set as a model parameter. 
		This is of comparable simplicity to our treatment, indeed both of our formalisms 
		require only two response functions. 
		It therefore should be possible to also fit this model for the full range of Fourier frequencies, 
		and comparison of our two different formalisms may provide information on the physical 
		origin of the continuum lags. 
		
		An interesting result we find here is that the low frequency lag spectrum 
		is predicted to dip dramatically at 
		$\sim 2$~keV (see top left of Fig.~\ref{fig:lag_fq}), which \citet{Uttley2011} see for GX 339-4. 
		They attribute it to fluctuations propagating in from the disc, 
		which could well be the case because of the shape of the covariance spectrum, 
		but it is interesting to note that we naturally expect this dip at $\sim 2$~keV 
		without any disc component.
		This is true only if we account for the non-linear effects, indicating 
		that ignoring them could bias the results.
		We note, however, that the introduction of a low energy 
		cut-off in the continuum spectrum of our model 
		could modify the lags at such low frequency. 
		
		Since the model is analytic, fits to data for many frequencies are 
		feasible without prohibitive computational expense.
		We fit hard state Cygnus X-1 data, and  
		find an unphysically low value for the disc inner radius in our best-fit. 
		Moreover the fit shows some residuals in the 
		iron line energy range of the real part of the 
		covariance spectrum for low frequencies.
		In the data, the line appears to be broad, both in the time-averaged spectrum 
		and for non-zero Fourier frequencies, implying a small disk inner radius. 
		However, the variability amplitude of the reflection signal drops off with frequency, 
		which in itself favours a fairly large disk inner radius \citep{Gilfanov2000}. 
		It is this tension that is behind the residuals seen in Fig. 8. 
		However, the transfer function we used to model reflection
		in this illustrative fit is over-simplified. 
		In particular, including light bending 
		should go some way to improving the fit as it
		both increases the light travel 
		time of reflected X-rays and focuses them to the inner regions of the disk. 
		In any case, the inclusion of light bending will dramatically 
		alter the model for our best fitting parameters, 
		and so our results from Section 5 should be 
		seen only as a proof of principle of the method, 
		with further insight delayed until all GR effects are included in the analysis.
		
		The value of the disk inner radius of Cygnus X-1 in the hard 
		state has long been a subject of debate in the literature. 
		For example, reflection modelling of the same NuSTAR 
		dataset gives values of $ r_{\rm in} $ ranging from $ \simeq 1.3 R_{\rm g} $ 
		\citep{Parker2015} to $ \simeq 13-20 R_{\rm g} $  \citep{Basak2017}, 
		depending on assumptions about the continuum. 
		\citet{Rapisarda2017} obtain a similar result for the inner radius $\sim 20~R_{\rm g}$  
		fitting a propagating fluctuations model to a hard state observation of Cygnus X-1, 
		although they did not account for the reverberation effects on the lags.
		We also note that the high resolution data used by \citet{Parker2015} reveal 
		absorption features around the iron line which would not be detectable with RXTE data. 
		Since the absorption properties of Cygnus X-1 depend on the binary orbital phase 
		\citep[likely due to the wind from the companion;][]{Grinberg2015}, it is not clear if such absorption 
		features affect our data.
		Here, we fixed the black hole mass in our fits but note that this can be left as a free parameter in future. 
		We also note that the same model can be used for AGN, 
		and is also sensitive to black hole mass in this case.
		
		Aside from including light bending, a number of other 
		improvements can be made to our method. 
		More realistic geometries than the lamppost model can be explored in future, 
		albeit with the trade-off of extra complexity increasing computational expense. 
		Our model is designed such that any transfer function can be ported into our formalism, 
		allowing flexibility. 
		It will also be fairly simple to include variations in cut-off energy and ionisation parameter. 
		As with power-law index variations considered here, 
		this can be done using a Taylor expansion, and so will not add prohibitively to computational cost.
		Finally, we have assumed that reflection is instantaneous, as is routinely assumed in the literature. 
		In reality, the reflected flux will take some finite time to increase when 
		the illuminating flux increases (on top of the light-crossing time). 
		\citet{Garcia2013b} showed, albeit with a very simple model, that the reflection spectrum 
		for a stellar-mass black hole should indeed respond very quickly ($\sim 1$~ns) 
		to a rise in the illuminating flux, but the response to a \textit{drop} in illuminating 
		flux could take as long as $\sim 1$~ms for a very low disc density. 
		The different timescales occur because the rise time depends on 
		photoionisation, but the fall time depends on recombination. 
		Therefore, this response time may be relevant for reverberation mapping.

\section{Conclusions}
\label{sec:conclusion}
		We developed a formalism for X-ray reverberation mapping of accreting 
		black holes that enables characterisation of the full range of cross-spectral 
		properties for a wide range of Fourier frequencies. 
		We empirically model the continuum lags that dominate at low Fourier 
		frequencies, and self-consistently account for the effect of the 
		continuum lags on the reverberation signal. 
		We point out some of previous inaccuracies in the literature 
		associated with fitting models to the observed energy dependent 
		phase lag and variability amplitude, and employ real and imaginary parts
		of the complex covariance to easily circumvent such problems. 
		As a proof of principle, we have fitted our model to an RXTE observation 
		of Cygnus X-1 in the hard state.
		We assume an on-axis lamppost geometry, 
		and obtain a fit with reasonable $ \chi^2 $,	
		 albeit with systematic residuals 
		around the iron line that would likely be improved by the 
		inclusion of light bending in the reflection model. 
		We also note that more realistic geometries will impact these results. 
		Although here we fixed the black hole mass in our fits to $14.8~M_{\odot}$ 
		\citep[following][]{Orosz2011}, we note that the model is sensitive to the 
		black hole mass. 
		This optimised model for X-ray reverberation mapping using the information
		of the cross variability for a wide range of Fourier 
		frequencies can therefore be used in future as a 
		new way to measure the mass of stellar-mass and supermassive black holes. 

\section*{Acknowledgements}
The authors would like to acknowledge the anonymous referee for the very helpful comments and suggestions.
G.M. acknowledges support from NWO.
A.I. acknowledges support from NWO Veni and the Royal Society.




\bibliographystyle{mnras}
\bibliography{library} 




\appendix
\section{Complex covariance analysis}
\label{app:cross-spectrum_analysis}
Analysing real and imaginary parts means that any linearity of the signal in the time 
domain is preserved in the frequency domain. 
This is not the case for amplitude and phase. 
Here we consider two simple explicit examples to illustrate the superiority of 
considering real and imaginary parts instead of amplitude and phase.
The first concerns the response matrix of the instrument and the second 
concerns scenarios when multiple variable spectral components are present. 
We assume unity coherence and infinite ensemble averaging, 
allowing us to drop angle the brackets notation.

\subsection{Instrument Response}
It is often assumed that the instrument response 
can be applied to the cross-spectral amplitude (or covariance/rms), giving
\begin{equation}
|G_{\rm o}\left(I,\nu\right)|=\int_{0}^{+\infty} R_{\rm t}\left(E,I\right)|G\left(E,\nu\right)| dE
\label{eq:telescope_resp}
\end{equation}
where $G_{\rm o}\left(I,\nu\right)$ is the observed complex covariance  
in the specific energy channel $I$ and  $R_{\rm t}\left(I,E\right)$ is the instrument response.
Here, $G\left(E,\nu\right)$ is in units of photons per second per ${\rm cm}^2$
per keV, and $G_{\rm o}$ is in units of counts per second.
However, this expression is not correct in general. The correct expression is 
\begin{equation}
G\left(I,\nu\right)=\int_{0}^{+\infty} R_{\rm t}\left(E,I\right)\,G\left(E,\nu\right) dE.
\label{eq:telescope_resp2}
\end{equation}
The amplitude of the complex covariance for channel $I$ is then 

\begin{equation}
|G_{\rm o}\left(I,\nu\right)| = \sqrt{
	\begin{aligned}
	&\left[\int_{0}^{+\infty} R_{\rm t}\left(E,I\right){\rm Re}\left[G\left(E,\nu\right)\right] dE\right]^2+\\
	& \left[\int_{0}^{+\infty} R_{\rm t}\left(E,I\right){\rm Im}\left[G\left(E,\nu\right)\right] dE\right]^2
	\end{aligned}	
}.
\label{eq:telescope_resp3}
\end{equation}
where ${\rm Re}$ and ${\rm Im}$ respectively denote the real and the imaginary part.
For $Re\left[G\left(E,\nu\right) \right] \neq 0$ and $Im\left[G\left(E,\nu\right)\right] \neq 0$, 
Eq.~\ref{eq:telescope_resp} is therefore incorrect. In the special case where 
$Re\left[G\left(E,\nu\right) \right] = 0$ or $Im\left[G\left(E,\nu\right)\right] = 0$, 
Eq.~\ref{eq:telescope_resp3} does reduce to Eq.~\ref{eq:telescope_resp}.

Since the observed phase lags are often fairly small, Eq.~\ref{eq:telescope_resp}
is, in practice, often fairly close to true. 
However, using our formalism of treating real and imaginary parts separately 
introduces no mathematical errors, and is no more difficult to implement than considering
the amplitude and the phase. It is clear the instrument response cannot be applied to the phase 
in a manner analogous to Eq.~\ref{eq:telescope_resp}.

\subsection{Multiple Components Fitting}
Consider a signal $s_1\left(E,t\right) = p\left(E\right)a\left(t\right)$ 
and another signal $s_2 = q\left(E\right)b\left(t\right)$, with Fourier transforms 
$p\left(E\right)A\left(\nu\right)$ and $q\left(E\right)B\left(\nu\right)$.
Using Eq.~\ref{eq:cross-spectrum} it follows from the linearity
of the Fourier transform that the cross-spectrum of 
$s_1\left(E,t\right) + s_2\left(E,t\right) $ is 
\begin{equation}
C\left(E,\nu\right)=  A\left(\nu\right)p\left(E\right)F^*\left(\nu\right) + B\left(\nu\right)q\left(E\right)F^*\left(\nu\right);
\label{eq:cross_appendix}
\end{equation}
i.e. simply the sum of the two individual cross-spectra. 
However, this linearity is lost for the amplitude and the phase lag.
Setting, for simplicity, the amplitude and phase of the reference band 
to $|F\left(E_{\rm r}\nu\right)|=1$ and $\phi_F\left(\nu\right)=0$ respectively, 
the amplitude and phase lag of the cross-spectrum are  
\begin{align}
|C\left(E,\nu\right)|= & \sqrt{p^2|A|^2+q^2|B|^2+2pq{\rm Re}\left[A\,B^*\right]}
\nonumber\\
\tan\phi_C\left(E,\nu\right)=& \left[\frac{p\,{\rm Im}\left[A\right]+
	q\,{\rm Im}\left[B\right]}{p\,{\rm Re}\left[A\right]+
	q\,{\rm Re}\left[B\right]} \right],
\label{eq:cross_ampli_phase}
\end{align}
In both expressions we have dropped the energy and frequency dependence of the 
components so as not to weigh the notation. 
We see that the expression of the amplitude has a cross term that would not appear 
if we simply added the amplitude of the two components. 
The exception is if the two components are in phase with each other, since in this case the 
amplitude of two vectors added in the complex plane is equal to the sum of the two amplitudes.  
\citet{Kotov2001} considered the case where the contribution of one component (e.g. $q(E)B(\nu)$) 
is small compared to the total spectrum ($q<<p$), in that case a binomial expansion gives 
\begin{align}
|C\left(E,\nu\right) |\simeq & p\, |A| \sqrt{1+2\frac{q\,Re\left[A\,B^*\right]}{p\,|A|^2}}\nonumber\\
\simeq & p\, |A|   + q\,\frac{Re\left[A\,B\right]}{|A|} = p |A| + q |B| \cos\phi_{AB},
\end{align}
where $\phi_{\rm AB}$ is the phase difference between $A$ and $B$.
Therefore, the contribution from the cross-terms in this case is small, but the normalisation 
of the second component is modified by a factor $\cos\phi_{AB}$.
From Eq.~\ref{eq:cross_ampli_phase}, it is clear that summing lags of two components 
is not appropriate.

\section{Response function}
\label{app:transfer_1}
Eq.~\ref{eq:refl_patch} in the text refers to emission from a specific 
patch of the accretion disc area. If the we consider flat space the area is 
expressed by $r\,dr\,d\phi$ and the total observed reflection spectrum 
varying in time is calculated by 
integrating the specific flux over the entire disc
\begin{equation}
R\left(E,t\right)=
\int_{r_{\textmd{in}}}^{r_{\textmd{out}}} \int_{0}^{2\pi} K(r)
g^3 A\left(t-\tau\right) \mathscr{R}\left(E/g\right) r\, dr \, d\phi ,
\label{eq:flux_tot}
\end{equation}
where we have not specified the $r$ and $\phi$ dependence 
of $g$ and $\tau$ for brevity.
As defined in the main text, $E$ is 
the observed energy and  $\mathscr{R}$ is the restframe 
reflection spectrum. 
The factor $K\left(r\right)$ accounts for geometrical dilution of radiation 
from the the point source incident on the disc patch, and is given by 
\begin{equation}
 K\left(r\right) = \varepsilon\left(r\right) \frac{\cos i }{D^2} = \frac{h\, \cos i }{\left(h^2+r^2\right)^{\nicefrac{3}{2}}\, D^2 }.
\label{eq:K}
\end{equation} 
The blue shift $g\left(r,\phi\right)$ is given by   
\begin{equation}
 g\left(r,\phi\right)=\frac{\sqrt{-g_{tt} -2g_{t\phi}\omega-g_{\phi\phi} \omega^2} }{1+\omega r \sin\phi\sin i },
\label{eq:g}
\end{equation} 
where $\omega = 1/\left(r^{3/2} + a \right)$ is the angular velocity 
in the dimensionless units, and we have again ignored light-bending 
\citep[see][]{Ingram2015,Ingram2017}. 
Finally light-crossing lag, $\tau\left(r,\phi\right)$, (i.e. the difference between the path of the light 
traveling directly from the point source to the observer and the light reflecting 
from the disc divided over the speed of light $\tau =l/c$) is given by 
\begin{equation}
\tau c= \sqrt{r^2+h^2}-r\sin i \cos\phi +h \cos i,
\label{eq:l}
\end{equation}
when light-bending is ignored.

\section{Linearization}
\label{app:linearization}
Taylor expanding Eq~\ref{eq:refl_patch} around $\beta = 0$ gives 
\begin{align}
dR \left(r,\phi|E,t\right) \simeq& A\left[t-\tau(r,\phi)\right]{\bigg[} \mathscr{R}\left(E/g(r,\phi)|\Gamma\right)-\beta\left(t-\tau(r,\phi)\right)\nonumber\\ 
&\frac{\partial\mathscr{R}}{\partial \Gamma}\left(E/g\left(r,\phi\right),\Gamma\right)  {\bigg]} 
 K(r)\, g^3\left(r,\phi\right) r\, {\rm d}r\, {\rm d}\phi.
\label{eq:linear_1}
\end{align}
Setting $B\left(t\right) \equiv A\left(t\right)\beta\left(t\right)$, the reflected flux 
integrated over the entire disc is
\begin{align}
R\left(E,t\right) \simeq& \int_{t' = 0}^{t'=\infty}\int_{\phi = 0}^{\phi = 2\pi}\int_{r = r_{\rm in}}^{r_{\rm out}} A\left(t'\right) \delta\left(t-\tau-t'\right)\nonumber\\ 
&\mathscr{R}\left(E/g\left(r,\phi\right)|\Gamma\right) K\left(r\right) g^3\left(r,\phi\right) r\, {\rm d}r \, {\rm d}\phi\, dt' - \nonumber\\
&\int_{t' = 0}^{t'=\infty}\int_{\phi = 0}^{\phi = 2\pi}\int_{r = r_{\rm in}}^{r_{\rm out}} A\left(t'\right) \delta\left(t-\tau-t'\right) \nonumber\\
& \frac{\partial\mathscr{R}}{\partial \Gamma}\left(E/g\left(r,\phi \right)|\Gamma\right) K\left(r\right) g^3\left(r,\phi\right) r\, {\rm d}r \, {\rm d}\phi\, dt'.
\label{eq:linear_2}
\end{align}
From the definition of a convolution 
\begin{equation}
A(t) \otimes z\left(E,t\right) \equiv \int_{0}^{\infty} z\left(E,t-t'\right) A(t')\, {\rm d}t',
\label{eq:convolutions}
\end{equation}
where $z\left(E,t\right)$ is an arbitrary function of energy and time.
We can simplify to 
\begin{equation}
R\left(E,t\right) =  A(t) \otimes w\left(E,t\right) - B(t)\otimes w_1\left(E,t\right),
\label{eq:linear_3}
\end{equation}
where $w\left(E,t\right)$ and $w_1\left(E,t\right)$ are defined in the main text. 
For the continuum, it is clear from Eq.~\ref{eq:Dlinearization} that  
\begin{equation}
D\left(E,t\right) \simeq \left[A\left(t\right) + B\left(t\right) \ln E \right] E^{-\Gamma} {\rm e}^{-E/E_{\rm cut}}.
\label{eq:linear_4}
\end{equation}
Summing direct and reflected components, and using the convolution theorem 
gives Eq.~\ref{eq:Four_cont+refl} in the main text.  
The transfer functions can be written analytically as 
\begin{align}
&W\left(E,\nu\right) = \int_{r_{\rm in}}^{r_{\rm out}} \int_{0}^{2\pi} {\rm e}^{i2\pi \nu \tau} K(r) g^3 \mathscr{R}\left(E/g|\Gamma\right) r \,{\rm d}r \, {\rm d}\phi \nonumber\\
&W_1\left(E,\nu\right) = \int_{r_{\rm in}}^{r_{\rm out}} \int_{0}^{2\pi} {\rm e}^{i2\pi \nu \tau} K(r) g^3 \frac{\partial\mathscr{R}}{\partial \Gamma}\left(E/g|\Gamma\right) r \,{\rm d}r \, {\rm d}\phi,
\end{align}
where we neglected the $r$ and $\phi$ dependency of $g$ and $\tau$ for brevity.
In our code, we first calculate transfer functions for $\mathscr{R} = \delta\left(E - 1{\rm keV}\right)$ and 
$\frac{\partial\mathscr{R}}{\partial\Gamma} = \delta\left(E - 1 {\rm keV}\right) $ and then perform 
convolution operations in the energy variable with $\mathscr{R}\left(E|\Gamma\right)$ and 
$\frac{\partial\mathscr{R}}{\partial\Gamma} \left(E|\Gamma\right)$
for obtaining $W$ and $W_1$ respectively.

\begin{figure}
	\includegraphics[width=8.5 cm]{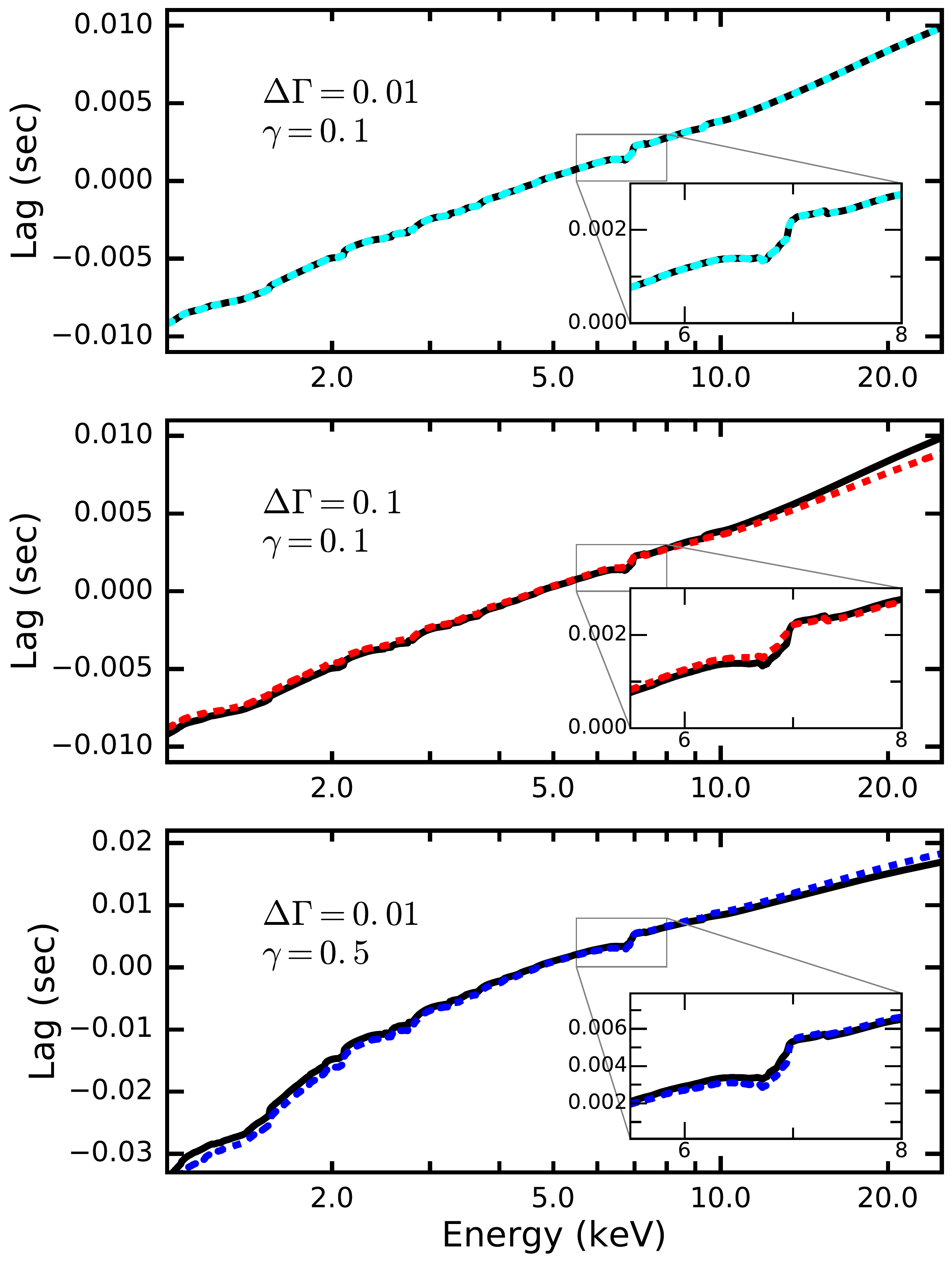}
	\caption{
		Predicted time lags as a function of energy for $2.5 {\rm Hz}$.
		For all the panels the solid curves are  the exact prediction (made in the time domain 
		without Taylor expansion) and the dashed curves are calculated using 
		the Eq.~\ref{eq:Dlinearization}~and~\ref{eq:Rlinearization} (first-order Taylor expansion). 
		For all the panels, $\phi_A=0\,{\rm rad}$ and $\phi_B = 0.2 \,{\rm rad}$. 
		In all panels we use the parameters: $\Gamma = 2$, $i=30\degree$, $r_{\textmd{in}}=10$, $r_{\textmd{out}}=100$,
		$h=10\,R_{\rm g}$, $a=0.998$, $M=10M_{\odot}$, $\log_{10}\xi = 3.1$, $A_{\rm Fe}=1$	}
	\label{fig:approximation}
\end{figure}

We checked under which conditions the linearisation of both the 
continuum and reflection expression is valid. 
Fig.~\ref{fig:approximation} shows the lag-energy spectrum at $2.5$ Hz 
calculated in the time domain without Taylor expanding either the continuum or the reflection 
expressions (solid lines) and analytically using first order linearisation (dashed lines).
We used a single frequency sine wave as input of the exact calculation of the lags, i.e.
$A\left(t\right) \propto 1+ \sin\left(2\pi\nu t - \phi_{\rm A}\right)$ and 
$B\left(t\right)\propto \sin\left(2\pi\nu t - \phi_{\rm B}\right)$.
The top panel (a) shows that for a reasonable choice of $\Delta \Gamma$  and 
a value of $\gamma$ similar to what we found from the fitting procedure the two 
curves match. However, if we decided to increase one of the two parameters 
(central and bottom panel) the Taylor expansion is less accurate.  

\section{Convolution of the theoretical model with the instrument response}
\label{app:fit_data}
Our model for the Fourier transform of the spectrum, $S(E,\nu)$, 
is given by Eq.~\ref{eq:Four_cont+refl} in the main text. 
Setting $Z(E,\nu)=E^{-\Gamma}{\rm e}^{-E/E_{\rm cut}} + W(E,\nu)$ and 
$Z_1(E,\nu)=E^{-\Gamma}{\rm e}^{-E/E_{\rm cut}} \ln E - W_1(E,\nu)$ 
gives the simplified form
\begin{equation}
S(E,\nu) = A(\nu) Z(E,\nu) + B(\nu) Z_1(E,\nu).
\end{equation}
Note that $Z$ and $Z_1$ are complex in general, due to the phase lags
introduced by the transfer functions $W$ and $W_1$. 
Convolving this around the instrument response (using the same procedure 
explained in Eq.~\ref{eq:telescope_resp2} in Appendix~\ref{app:cross-spectrum_analysis})
gives the Fourier transform of the observed spectrum
\begin{equation}
S(I,\nu) = A(\nu) Z(I,\nu) + B(\nu) Z_1(I,\nu),
\label{eq:S_channels}
\end{equation}
where $I$ indicates the $I^{\rm th}$ energy channel.
The Fourier transform of reference band flux is $F(\nu)$, and therefore
our model for the complex covariance of the $I^{\rm th}$ energy
channel is
\begin{equation}
G(I,\nu) = \frac{ A(\nu) F^*(\nu) Z(I,\nu) + B(\nu) F^*(\nu)
	Z_1(I,\nu) }{ |F(\nu)| }.
\end{equation}
To simplify further, we can simply set
\begin{align}
\alpha(\nu) {\rm e}^{i \phi_A(\nu)} &=
\frac{A(\nu)F^*(\nu)}{|F(\nu)|} \label{eqn:alpha}
\\
\alpha(\nu) \gamma(\nu) {\rm e}^{i \phi_B(\nu)} &=
\frac{B(\nu)F^*(\nu)}{|F(\nu)|}
\label{eq:gamma}
\end{align}
to get
\begin{equation}
G(I,\nu) = \alpha(\nu) \left[ {\rm e}^{i \phi_A(\nu)} Z(I,\nu) +
\gamma(\nu) {\rm e}^{i \phi_B(\nu)} Z_1(I,\nu) \right].
\end{equation}
This expression is identical to Eq.~\ref{eq:complex_covariance_channels}
that we derived in the main text convolving the final complex covariance model  
(Eq.~\ref{eq:complex_covariance_model}) with the instrument response.

\section{Reference band}
\label{app:model_parameter}

Following \citet{Revnivtsev1999} and \citet{Gilfanov2000}, 
who analysed the observations considered here, we choose channels $ 5-7 $ ($ 2.84 - 3.74 $ keV) 
as our reference band. This choice is motivated by these channels having high 
count rates and being disjunct from the `science' channels (and thus ensuring statistical 
independence between the reference band and all of the science channels).
Moreover the $ 2.84 - 3.74 $ keV range does not contain any of 
the reflection features that we are interested in observing. 
When the complex covariance is calculated (for the data) in the reference  
band energy range, the imaginary part is identically zero because the complex covariance 
of the reference band with itself has zero phase by definition.
This could, in principle, impose an additional constraint that 
involves all the continuum parameters of the model:
\begin{equation}
\sum_{I=I_1}^{I_2} {\rm Im} \left[ G(I,\nu) \right] = 0,
\label{eq:extra_condition}
\end{equation}
where $I$ indicates the $I^{\rm th}$ energy channel, $I_1$ to $I_2$ 
set the reference band channel range and $ G(I,\nu) $ is defined in 
Eq.~\ref{eq:telescope_resp2}. 
In our case it is not possible to apply this extra condition because 
the available instrument response $ R'_{\rm t}(E, I) $ is poorly calibrated 
in the energy range of the reference band, while the relation between 
$ G(E, \nu) $ and $ G(I, \nu) $ is of course always defined by the {\it true} 
instrument response $ R_{\rm t}(E, I) $. 
A mathematical way of looking at it is that we construct a model for the 
Fourier frequency dependent spectrum $ S(E, \nu) $, which we cross with 
the Fourier transform of a model reference band, $ F(\nu) $, to get the 
predicted complex covariance
\begin{equation}
G(E,\nu)  = \frac{S(E,\nu) F^*(\nu)}{|F(\nu)|}   .
\label{eq:complex_cavariance_app}
\end{equation}
If we had a response matrix that were well calibrated for the reference 
band energy channels, we could self-consistently calculate 
the reference band Fourier transform from our spectral model
\begin{equation}
F(\nu)=\int_0^\infty \sum_{I_1}^{I_2} R_t(E, I) S(E, \nu) dE .
\label{eq:reference_band_app}
\end{equation}
However, we do not know the true response $ R_{\rm t}(E,I) $, and 
so we simply leave the phase angle of the unity magnitude complex 
number $F(\nu)/|F(\nu)|$ as a model parameter that gets swallowed 
up into the definition of the parameters $\phi_A(\nu)$ and $\phi_B(\nu)$. 
In order to calculate $ G(I,\nu) $ to fit to the data, we convolve $ G(E,\nu) $ 
with the {\it available} response $ R'_t(E,I) $.  
This response is actually well calibrated in the range where we fit it to the data, 
so assuming the model is correct we obtain the correct values for the fit parameters 
and hence correctly recover $ G(E, \nu) $. 
If we could convolve this $ G(E, \nu) $ with the true response $ R_{\rm t}(E, I) $, 
the $ G(I, \nu) $ obtained from this operation would satisfy condition~\ref{eq:extra_condition}.
However, if we convolve $ G(E,\nu) $ with the available response $ R'_{\rm t}(E,I) $ 
which is poorly calibrated and hence differs from $ R_{\rm t}(E, I) $ between $ I_1 $ and $ I_2 $, 
we obtain
\begin{equation}
\frac{\int_0^\infty \sum_{I_1}^{I_2} R'_{\rm t}(E, I) S(E, \nu) dE \,\,F^*(\nu) }{|F(\nu)|} = \frac{F'(\nu)F^*(\nu)}{|F(\nu)|},
\end{equation}
whose imaginary part is not identically zero. 
For this reason, using the available response, condition~\ref{eq:extra_condition} 
is not expected to apply to our chosen reference band. 
Indeed, when we perform this experiment this turns out to be the case, 
although the discrepancy is not very large, indicating the calibration, while poor, 
is still passable even in the energy range of the reference band. 
It is worth noting that the poor calibration of the reference band does not affect 
our measurements of $ G(E,\nu) $. 
Since we divide through by the modulus of $ F(\nu) $, 
all that is affected is the phase of our reference band model $ F(\nu) $. 
At any given frequency this only introduces a phase offset that is the same at each energy, 
not affecting the physically meaningful phase differences between energy bands.  
In the absence of a physical time series model, however, 
the phase differences between Fourier frequencies in a given 
energy band can not be used to constrain the physical parameters.


\bsp	
\label{lastpage}
\end{document}